**Revealing Hidden Vibrational-Polariton Interactions by 2D IR Spectroscopy**

Bo Xiang[2], Raphael F. Ribeiro[1], Adam D. Dunkelberger[3], Jiaxi Wang[1], Yingmin Li[2], Blake. S. Simpkins[3], Jeffrey C. Owrutsky[3], Joel Yuen-Zhou[1], Wei Xiong[1,2]

[1] Department of Chemistry and Biochemistry, University of California, San Diego, La Jolla, CA

[2] Materials Science and Engineering Program, University of California, San Diego, La Jolla, CA

[3] Chemistry Division, Naval Research Laboratory, Washington, District of Columbia

*Abstract* We report the first experimental two-dimensional infrared (2D IR) spectra of novel molecular photonic excitations – vibrational-polaritons. The application of advanced 2D IR spectroscopy onto novel vibrational-polariton challenges and advances our understanding in both fields. From spectroscopy aspect, 2D IR spectra of polaritons differs drastically from free uncoupled molecules; from vibrational-polariton aspects, 2D IR uniquely resolves hybrid light-matter polariton excitations and unexpected dark states in a state-selective manner, and revealed hidden interactions between them. Moreover, 2D IR signals highlight the role of vibrational anharmonicities in generating non-linear signals. To further advance our knowledge on 2D IR of vibrational polaritons, we develop a new quantum-mechanical model incorporating the effects of both nuclear and electrical anharmonicities on vibrational-polaritons and their 2D IR signals. This work reveals polariton physics that is difficult or impossible to probe with traditional linear spectroscopy and lays the foundation for investigating new non-linear optics and chemistry of molecular vibrational-polaritons.

Molecular polaritons have been explored as a novel tool to harness confined electromagnetic fields to control chemical reactivity through modified vibrational dynamics, tailored potential energy landscapes, and long-range mesoscopic energy transfer[1–8]. Polaritons can be described as delocalized quantum superpositions of material and electromagnetic modes resulting from strong coupling between them[9,10]. The emergence of these hybridized states is similar in nature to the formation of molecular orbitals arising from the overlap of atomic states[11], Fermi resonances from couplings between molecular vibrations[12], or delocalized states in H- and J-aggregates from dipolar interactions between localized excitons[13]. Vibrational-polaritons arise from the strong interaction between a confined photonic mode and vibrational excitations across the molecular sample (Fig.1 a-c)[9]. Such coupling relies on the cumulative material oscillator strength that allows for efficient vibration-cavity hybridization resulting in delocalized polaritons that extend across a macroscopic number of molecules in the entire photonic mode (Fig.1b). This coherence may enable control of molecular processes[9] and, in fact, impressive developments in exciton-polariton research have already been achieved and include room-temperature, tabletop realizations of superfluids[14], polariton lasing[15], and Bose-Einstein condensates[10,16–18]. We anticipate equally exciting developments in the molecular sciences with the advent of vibrational-polaritons.

Recently, the steady-state response of vibrational-polaritons has been probed in polymers, neat liquids, and solutions[1,3,4,6] and revealed enhanced Raman scattering[2] and modified kinetics of chemical reactions, even in the absence of external pumping[19]. The first time-resolved spectroscopic study of these systems showed vibrational-polariton dynamics that were significantly faster than that of the uncoupled molecule and highly sensitive to the light-matter composition of the polaritons (*i.e.*, the Hopfield coefficients)[19]. However, important aspects of vibrational-polaritons are still not understood. For instance, theoretical models show that there are many dark states which do not couple to the cavity mode in comparison to the few bright states that do[9,20,21]. Despite their "dark" character, strong transient features have been attributed to this reservoir of dark states[22], raising important questions about how these states are populated and interact with the optical cavity mode[19]. Resolving state-selective excitation and population transfer, unique capabilities afforded by two-dimensional infrared (2D IR) spectroscopy, will be key to understanding and utilizing polaritons for chemical control. We note here an additional distinction between vibrational- and exciton-polaritons. Exciton-polaritons are commonly treated as composite bosons where the biexciton binding energy is ignored and the main source of nonlinearities is the so-called "phase space filling" effect which arises due to Pauli exclusion[23,24]. In contrast, vibrational excitations do not experience Pauli repulsion but feature nuclear and electrical anharmonicities. It is thus important to understand how these fundamental differences influence the energy landscape of vibrational-polaritons.

In the following, we investigate the aforementioned questions by ultrafast 2D IR spectroscopy[25], the first measurement of its kind for vibrational-polaritons. 2D IR spectroscopy probes the nonlinear optical response of IR-active modes, with the ability to resolve the pump excitation frequency, which makes it uniquely suited to investigate interactions between states key to polariton dynamics. The 2D IR data reveal unexpected interactions between dark states and vibrational-polaritons as well as direct excitation of dark states. Furthermore, the mere existence of vibrational-polariton pump-probe[19] and 2D IR signals is direct evidences of deviations from the harmonic polariton model and motivates our development and presentation of a theoretical treatment of nonlinear polariton response incorporating vibrational anharmonicities.

**Results**

The sample is composed of a nearly saturated $W(CO)_6$ solution in a microcavity consisting of two dielectric stack mirrors separated by a 25 $\mu$m spacer (see Methods). In the strong coupling regime, upper and lower polariton branches (termed UP and LP) are formed upon hybridization of vibrational and cavity modes. By adjusting the tilt angle, $\theta$ (Fig.1a), the cavity resonance can be tuned across the vibrational transition, thus revealing a dispersive anti-crossing and allowing control of the polariton light-matter composition (Fig.1d).[1,3,4,9,18] When the cavity-vibration detuning is zero (*i.e.*, frequency of both is 1983 cm$^{-1}$), LP and UP have equal photonic and delocalized-vibration characters (Fig. 1c). The energy difference between the polariton branches at this condition (~37 cm$^{-1}$) corresponds to the vacuum Rabi splitting. Blue-tuning the cavity probes an LP transition with stronger vibrational character while the UP has a higher photonic component. Red-tuning has the opposite effect.

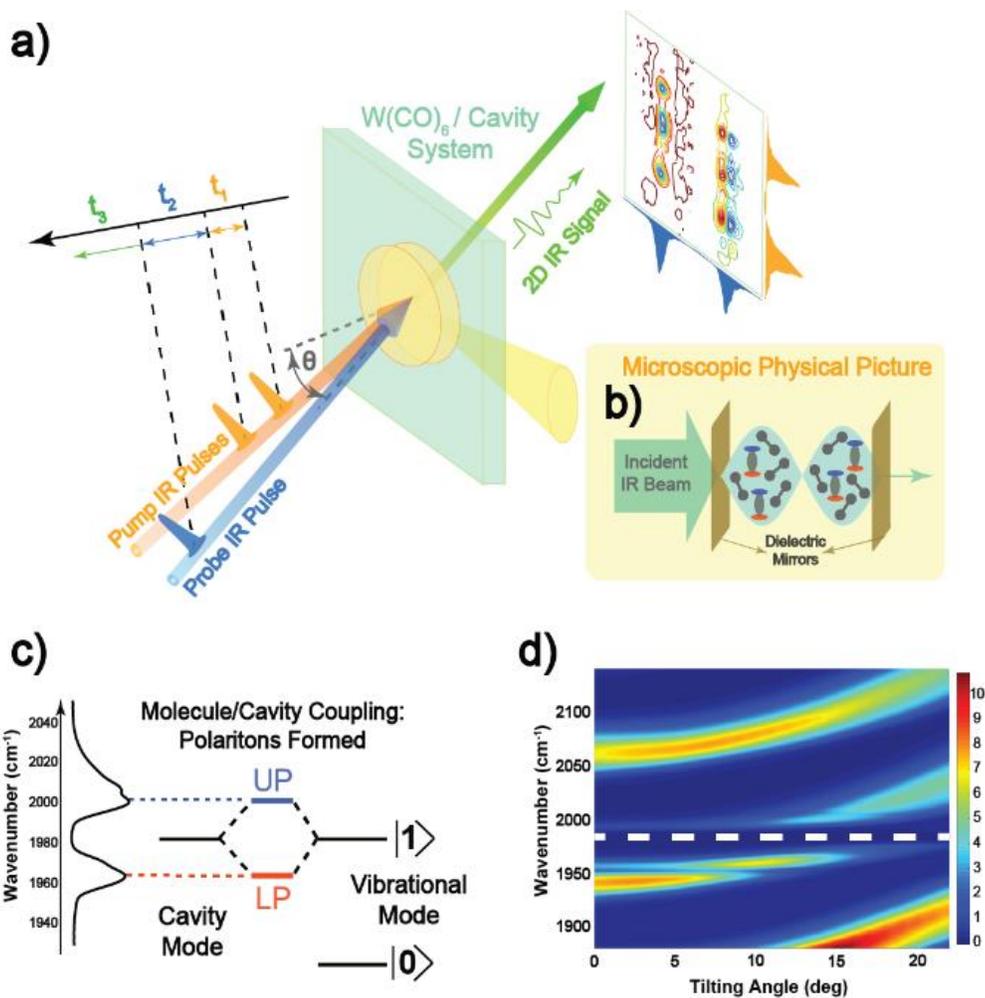

Figure 1. (a) Scheme of vibrational polariton 2D IR spectroscopy setup (pump and probe IR incident beams are symmetric with respect to the normal plane with the same tilting angle, θ. The coherences are characterized by scanning $t_1$ and $t_3$, and the resulting time-domain interferograms are Fourier transformed to obtain 2D IR spectra. The other delay time, $t_2$, is equivalent to the delay between the pump and probe pulses in pump-probe spectroscopy. (b) Illustration of the microscopic physics of molecules inside of a microcavity where 'pure grey modes' correspond to the vibrational modes that are not strongly coupled to the cavity and the rest refers to the strongly coupled modes that contribute to UP (blue), LP (red) and dark reservoir modes (grey) formation. (c) Formation of vibrational polaritons by strongly coupled molecular vibration and cavity modes. Left, vibrational polariton FTIR spectrum of W(CO)$_6$. (d) Dispersion of IR transmission, as a function of the tilting angle, θ, of a microcavity filled with a nearly saturated W(CO)$_6$ in hexane solution, white dotted line indicating the vibrational frequency of W(CO)$_6$.

We first confirm that the pump-probe response of molecules both inside and outside the cavity agree with published results[19,26,27]. The transient spectrum for uncoupled $W(CO)_6$ (Fig. 2a) is well-understood. The negative feature at 1983 $cm^{-1}$ corresponds to ground state bleach and stimulated emission while positive features indicate absorption in newly populated excited states. Importantly, for this uncoupled system, only the peak amplitudes vary with excited-state population, *i.e.*, the spectral position of each transient feature remains fixed at the frequency corresponding to the excited or bleached state irrespective of the pump-induced transient population. Notable qualitative differences are observed between the bare molecule pump-probe response (Fig.2a) and that of the strongly-coupled system (Fig.2b). As further explained below, the distinctive features in the polaritonic spectra do not necessarily correspond to transient population changes of states. Instead, they are due to the interaction of cavity photons with the matter polarization originating from excited-state absorption, ground-state bleach, and stimulated emission of reservoir modes.

## *2D IR Spectra of Vibrational Polaritons*

2D IR spectroscopy reveals coherences between states using a three IR pulse sequence (Fig. 1a). We focus on results obtained at $t_2$ = 25 ps, where spectral oscillations have ceased, and after which only incoherent polariton-polariton and polariton-dark state interactions are significant. There are sharp contrasts between the 2D IR spectra of uncoupled $W(CO)_6$ (Fig.2c shows ground state bleach/stimulated emission and excited-state absorption at the frequencies of those transitions) and cavity-coupled $W(CO)_6$ (Fig. 2d shows strong excited-state absorption features extending along the $\omega_1$ axis and weaker features near the UP region). We have eliminated the possibility of a cavity-induced spectral filtering effect contributing to the 2D IR spectra (SM section 2e).

Details of the cavity-coupled response are more clearly seen in Fig. 3 where we have rescaled each region for visibility. Figure 3 shows many features along the diagonal, where the system is pumped and probed at the same frequency, as well as cross peaks, indicating coupling or energy transfer between modes. Along the diagonal, we observe a large absorptive feature in the LP region ($\omega_1=\omega_3=\omega_{LP}$) (Fig.3c), and a derivative-like modulation in the UP region ($\omega_1=\omega_3=\omega_{UP}$) (Fig.3d). Additionally, "cross peaks" appear when exciting either polariton mode[28]. Specifically, when the UP is excited, a strong feature (Fig. 3a) identical to that of Fig. 3c is observed when the system is probed at the LP frequency ($\omega_1=\omega_{UP}$, $\omega_3=\omega_{LP}$), and the derivative lineshape observed in Fig.3d also appears when the LP is excited ($\omega_1=\omega_{LP}$, $\omega_3=\omega_{UP}$) (Fig.3f). Calculations based on Fresnel expressions[19] (also see SM section 3) associate the large, absorptive feature located at the LP region (~1960 cm$^{-1}$) with the coupling of the cavity mode to v = 1 to 2 polarization (due to transient v = 1 reservoir population), while the derivative lineshape in the UP region and weak negative responses at the edges of the LP region are ascribed to an effective Rabi splitting contraction induced by reduction of ground state (v = 0) population. Below, a microscopic quantum-mechanical model will be presented which supports these interpretations (also see SM sections 4).

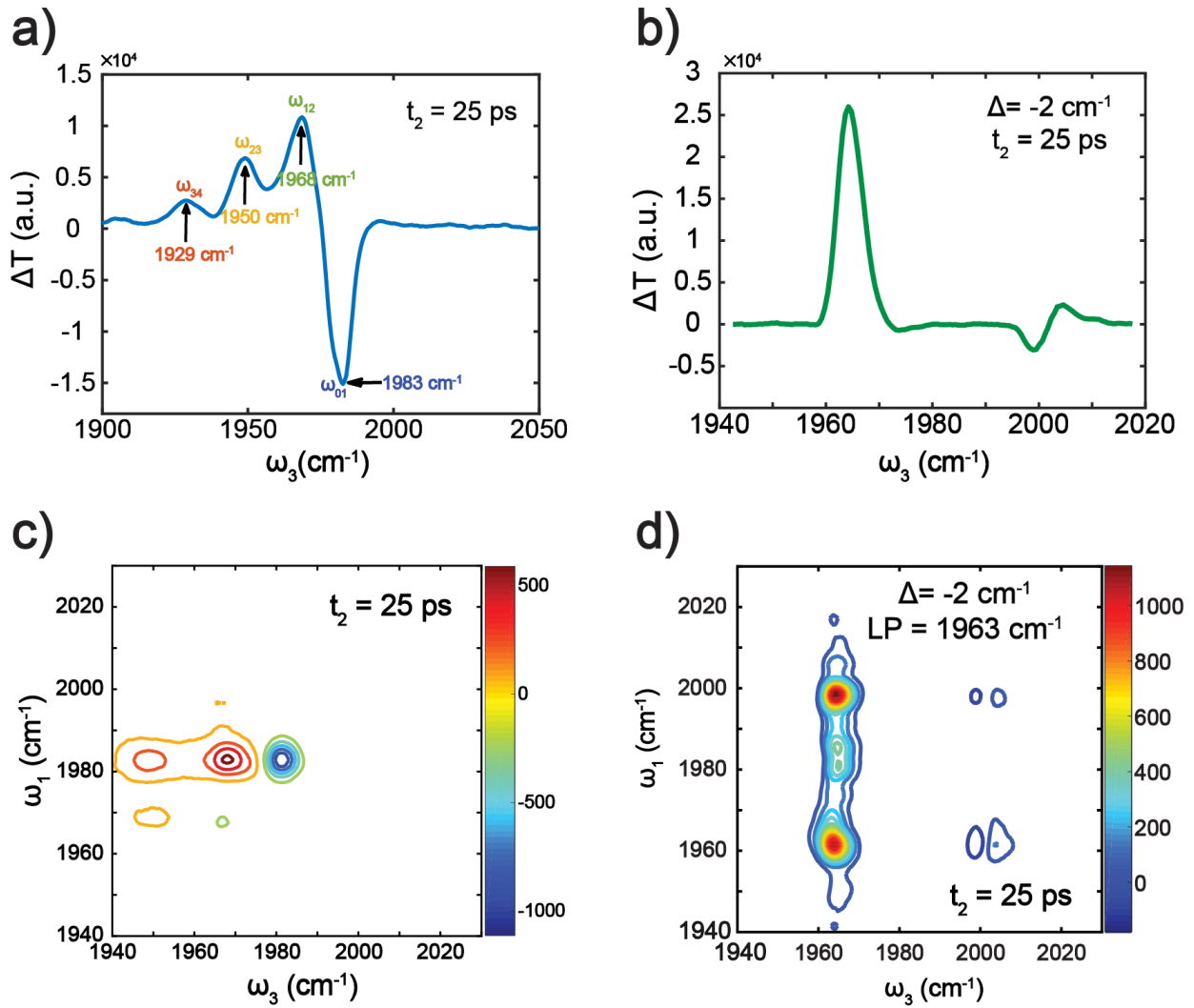

Figure 2. Pump-probe spectra of (a) W(CO)$_6$ outside of the cavity. (b) W(CO)$_6$/cavity-vibrational polariton system with IR cavity mode detuning ($\Delta$) of -2 cm$^{-1}$. By setting $t_1$ = 0, we can collect traditional pump-probe spectra. (c) 2D IR spectra of W(CO)$_6$ outside of cavity. (d) 2D IR spectra of W(CO)$_6$/cavity polariton system with IR cavity mode detuning of -2 cm$^{-1}$. 2D IR are plotted against $\omega_1$ (essentially the excitation frequency) and $\omega_3$ (the probe frequency), the coherence frequencies during $t_1$ and $t_3$, respectively and with $t_2$ = 25 ps.

*Unexpected 2D IR peaks reveal dark state populations*

A third set of features requires discussion. When the system is excited at the uncoupled $\omega_{01}$ frequency, and probed at the LP and UP frequencies, (Fig. 3b and e), peaks are observed, consistent with the presence

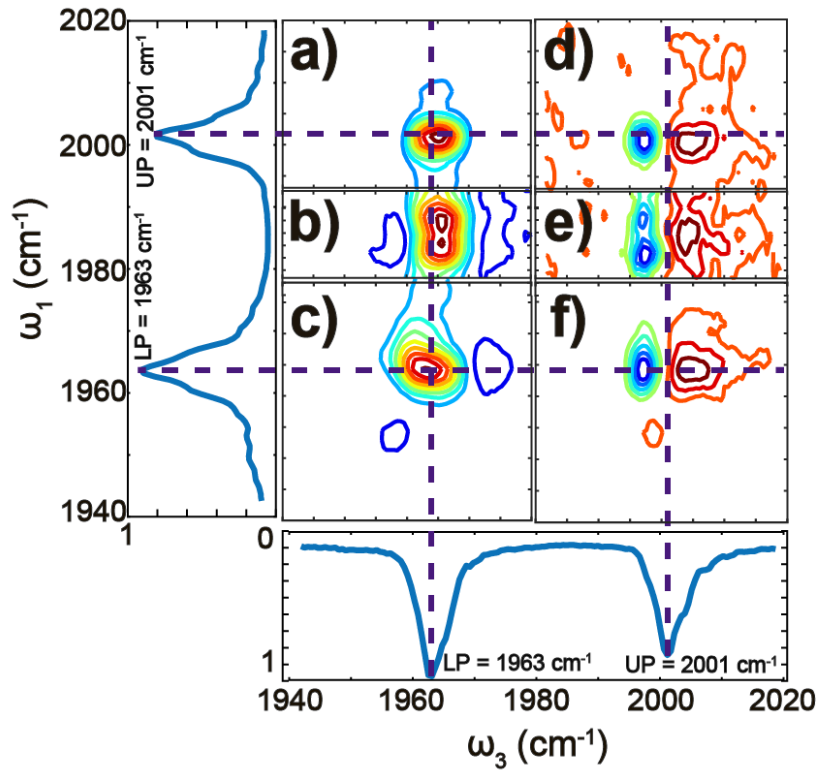

Figure 3. 2D IR spectrum of W(CO)$_6$/cavity polariton system at 25 ps delay with -2 cm$^{-1}$ detuning. Each spectral region is scaled to its own intensity maximum and minimum. Spectra of the pump ($\omega_1$) and probe ($\omega_3$) pulses are shown on their respective axes. Color map: red is positive, blue is negative

of v = 1 reservoir population. While these features are similar to those discussed above, their presence when exciting at the uncoupled $\omega_{01}$ frequency indicates direct excitation of reservoir states of $W(CO)_6$ despite the "dark" nature of these states. This direct excitation of reservoir modes is comparatively weak (see unscaled data of Fig.2d). We believe direct excitation of dark modes occurs by way of non-unity cavity reflections and disorder in the molecular system[20]. This phenomenon is analogous to the observed direct excitation of bare excitons in systems with exciton-polaritons.[23]

The features in Fig. 3b and 3e have not been observed in any polariton coherent 2D spectroscopy and indicate an interaction between the so-called dark modes and bright polaritons – when one mode (dark) is excited, it influences the optical response of the others (polaritons)[24]. A similar interaction, but in the opposite order (exciting polaritons and influencing dark modes), is also evident in the 2D IR spectra. The large absorptive features in Fig. 3a and c are due to reservoir excited state population and occur after selective excitation of either polariton. The fact that response is qualitatively similar when exciting at the LP, UP, or even the dark reservoir frequency, indicates that there are polariton-reservoir interactions that lead polaritonic excitations (either LP or UP) to transfer into reservoir states possibly via phonon scattering[29,30].

The existence of reservoir mode peaks imposes difficulty in disentangling certain spectral signatures. In principle, the large positive signals observed near the LP (Figs. 3a, b and c) are a mixture of two contributions: nonlinear polariton optical response and excited-state absorption from the reservoir hot bands[19]. In contrast, the 2D spectra that probe UP states are easier to understand (Figs. 3d, e and f), and reveal interactions between various states. For example, the derivative-like feature in Figs.3e and 3f indicates that by populating either the LP or dark reservoir modes, the UP frequency is modulated. Such insights are obfuscated when probing near the LP due to its near-resonance with the reservoir excited-state absorption.

**Cavity-Detuning-Dependent Response**

Important insights arise from examining the system response as a function of cavity detuning. The data in Fig. 4 correspond to spectral cuts along the probe frequency axis ($\omega_3$) with the pump frequency ($\omega_1$) fixed at the reservoir mode ($\omega_{01}$), LP, and UP frequencies ($\omega_{LP}$ and $\omega_{UP}$) (4a, b, and c, respectively). The cavity is detuned by rotating the sample as described in Fig.1a. When exciting at $\omega_1$ = 1983 cm$^{-1}$ (dark state, Fig.4a), a large positive feature is seen near the LP region; it is maximal when the cavity is tuned such that the LP frequency is 1968 cm$^{-1}$. On the other hand, while the spectral cuts at LP and UP frequencies (Fig.4b and c) show a similar feature, it is maximized when LP is tuned to 1963 cm$^{-1}$. Identifying this striking contrast relies on the capability of 2D IR for state-selective excitation which reveals aspects of the response that cannot be discerned from integrated pump-probe signals (Fig. S13).

As discussed above, the large feature in the LP region can be attributed to the existence of an excited-state population in the $v_1$ reservoir states (i.e., the first vibrational excited state). Therefore, the detuning-dependent intensity of this peak indicates either (i) a varying sensitivity to the reservoir population or (ii) a detuning-dependent efficiency in generating reservoir population. First, we examine Fig. 4a where reservoir modes are excited directly. In this case, the signal is maximized when the LP resonance is tuned to the $\omega_{12}$ band (~1968 cm$^{-1}$). This behavior is analogous to a cavity-enhanced optical response where the excited-state population of reservoir states is most sensitively probed when the polariton resonance is coincident with the transition being probed[31,32]. In other words, the reservoir population may not be influenced much by tuning but our ability to see it is. However, when exciting the polaritons (Fig.4b and c), this feature is maximized when the LP frequency is 1963 cm$^{-1}$. This frequency does not align with the reservoir $\omega_{12}$ transition, but it is near the condition of zero detuning between the cavity and the vibration.

Under these excitation conditions, polaritons are first excited and then decay into the $v_1$ reservoir causing the nonlinear optical response near the LP. We propose that the generation of polariton excitations, which subsequently decay into reservoir population, is maximized when the initial polariton absorption is maximized (*i.e.*, at zero detuning; see SM section 3). We note that the calculation of cavity detuning is highly sensitive to the calibration of the spectrograph and the current values are determined based on both frequencies and intensities of the LP and UP. Nevertheless, these results indicate that even after a relatively long delay time of 25 ps (relative to the cavity lifetime, which is < 5 ps), the resultant excited-state population can be manipulated via the excitation frequency (polariton vs reservoir excitation) or angle (the relative photon-vibration character of the polariton), which may prove useful in future implementation of cavity-modified molecular excitations.

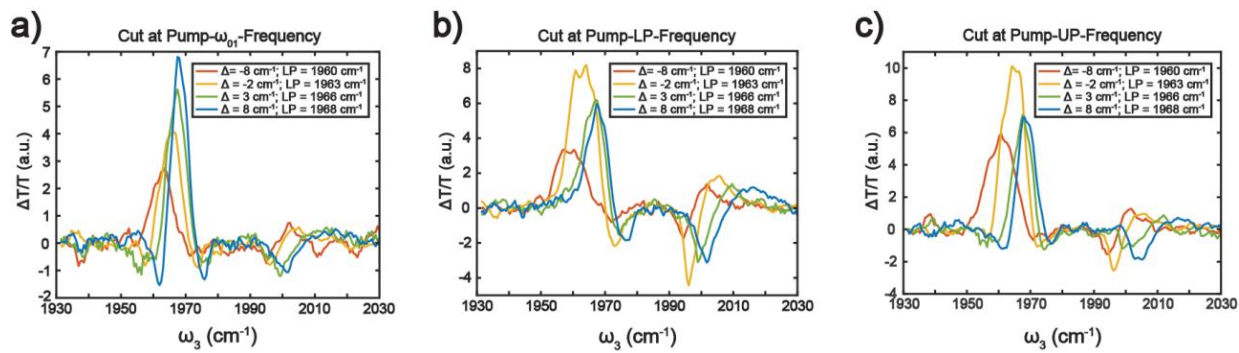

Figure 4. Scaled spectral cuts at pump frequency of (a) $\omega_{01}$, (b) LP, and (c) UP with different detuning (Δ).

### *Origin of 2D IR spectra of vibrational-polaritons: deviations from the free harmonic boson*

Even though we have shown that 2D IR spectra reveal dark states and their interactions with polaritons, the origin of the 2D IR spectra of vibrational-polaritons is an interesting topic in its own right. Given previous emphasis on their linear response, polaritons have been primarily described within the free boson picture, featuring harmonic oscillator dynamics[24,33,34]. However, it is well-known that harmonic systems exhibit vanishing nonlinear optical response[35]. Thus, the observation of vibrational-polariton pump-probe[19] and 2D IR signals necessarily indicates that this free bosonic picture does not generally hold, and that anharmonicities in the system must exist to give rise to nonlinear spectra. In inorganic semiconductor exciton-polaritons, effective anharmonicities are induced by Coulomb scattering and the so-called "phase space filling" mechanism due to Pauli exclusion of electrons and holes[24,36]. For vibrations, nuclear and electrical (or dipolar) anharmonicity[37,38] of the C-O stretch causes deviations from harmonic behavior. As the level of excitation in the system increases, v = 1 to 2 transitions become coupled to the cavity, in addition to the fundamental transition, thus influencing polariton formation and dispersion. Both the red-shift of $\omega_{12}$ compared to $\omega_{01}$ (nuclear anharmonicity) and the reduced oscillator strength $\mu_{12}$ compared to $\sqrt{2}\mu_{01}$ (electrical anharmonicity) lead to a red-shift of the UP (*i.e.*, contraction of the vacuum Rabi splitting leading to derivative-like lineshape). Classical expressions for transmission through a cavity capture this effect well with respect to reservoir population, but here we develop a full quantum-mechanical treatment to provide a microscopic perspective and shed light on the conditions of validity of the classical approximation.

The details of our approximate quantum-mechanical (QM) model are described in SM section 4. The Hamiltonian includes the effects of both nuclear and electrical anharmonicity, which may be understood to induce self- and cross-interactions between the LP and UP, as well as between polaritons and dark states. With either the classical expression or the QM model, we can estimate the degree of reservoir excitation by comparing the modeled and experimentally determined shift of the UP. From the classical model, we estimate that 5% of the molecules are excited at $t_2$ = 25 ps, while the QM model suggests 5-7.5%. Both estimates are reasonable for the pulse energies used. The QM model captures the main features of the vibrational-polariton (Fig.5) pump-probe signal, including excited-state absorption, stimulated emission, and ground-state bleach, as well as the order of their corresponding intensities. We only present a comparison of experimental and theoretical pump-probe signals because simulation of the complete 2D spectrum requires a detailed theory of nonlinear multi-polariton dynamics including many-body interactions between dark-states and polaritons, which we are currently developing.

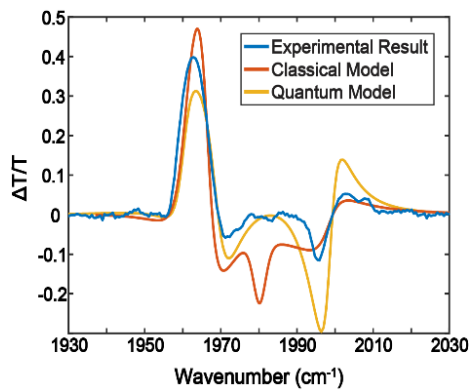

Figure 5. Comparison between the experimental pump-probe signal and the classical and quantum models. Reasonable qualitative agreement is achieved.

**Discussion**

While the assumption of transient dark state population seems sufficient to capture several of the described 2D IR features (*e.g.*, the derivative lineshape at UP and large absorptive feature at LP), the differences of 2D IR signals taken at $\omega_1$ = LP, UP relative to those at $\omega_1$ = $\omega_{01}$ (Fig. 4) indicate that the time-dependent population of reservoir modes from polariton relaxation is surprisingly sensitive to cavity detuning. The quantum description presented here reproduces all of the characteristic features of the transient response at longer times and a full treatment which includes the dynamic coherence and population transfer between polaritons and dark states is under development.

The multidimensional spectroscopy of the cavity-coupled C-O stretch of $W(CO)_6$ reveals direct excitation of reservoir states (despite their weak linear response) and unambiguous evidence of interactions between reservoir and polariton modes. The visibility of direct dark mode excitation in 2D IR signals arises from the detection axis $\omega_1$, where vibrational coherences are detected through free-induction-decay (FID) along $t_1$ of the emitted signal, instead of directly measuring the emitted optical responses, which can be weakened by reabsorption. These results highlight the value of 2D spectroscopy which follows quantum coherences in the time domain and resolves them in an additional frequency domain revealing complex physical processes, as demonstrated in many conventional condensed-phase systems[39–43]. These features of 2D spectroscopy can be extended by performing polarization resolved measurements, quantum process tomography[44], or even 3D spectroscopy[45–47] for tracking emergent two-point correlations between vibrational-polaritons and uncoupled vibrations. Many novel liquid-phase molecular systems deserve attention, including heterogeneous systems, intrinsically coupled[37], and decoupled molecular vibrations, and molecular systems that undergo isomerization[48]. In addition, many novel nanoplasmonic optics can be designed and used to enhance local electric field and hence the coupling strength[49]. Given its ability to access precise details of excited state dynamics, we expect multidimensional spectroscopy to be a workhorse in the study of molecular polaritons.

*Methods*

**Molecule-polariton Preparation.** The $W(CO)_6$ (Sigma-Aldrich) /cavity system is prepared in IR spectral cell (Harrick) containing two dielectric $CaF_2$ mirrors separated by a 25 μm spacer and filled with nearly saturated $W(CO)_6$/hexane solution. The dielectric mirror has a ~96% reflectivity. Because the Rabi splitting (37 $cm^{-1}$) is larger than the full-width-at-half-max of both cavity (~11 $cm^{-1}$) and $W(CO)_6$ vibrational (~3 $cm^{-1}$) modes, the strong coupling criteria is satisfied. The transmission spectra of both pump and probe polaritons are measured along with 2D IR spectra.

**Angular-resolved 2D IR spectrometer.** The spectrometer follows the pulse shaper enabled pump probe design[50], and a rotational stage is added to control beam incidence angle. In the 2D IR spectrometer, three IR pulses interact with a sample sequentially to create two vibrational coherences. The first coherence is characterized by scanning $t_1$. The second coherence introduces a macroscopic polarization which subsequently emits a third order IR signal, which is self-heterodyned and detected in frequency domain. To display 2D IR spectra, the FID in $t_1$ is numerically Fourier transformed.

**Theoretical model.** As described in SM, the QM model utilizes a Hamiltonian including a single-cavity mode interacting weakly with external vacuum electromagnetic modes, and with the polarization generated by *N* independent anharmonic molecular vibrations, which are themselves weakly-coupled to low-energy bath modes. We utilize input-output theory to estimate the probe transmission in the presence and absence of a transient excited-state population of molecular vibrations.

***Acknowledgements.*** B.X., J.W., Y.L., W.X. are supported by AFOSR Young Investigator Program Award, FA9550-17-1-0094. B.X. thanks Roger Tsien Fellowship from UCSD Department of Chemistry and Biochemistry. B.S.S., A.D.D., J.C.O. are supported by the Office of Naval Research through the Institute for Nanoscience at the Naval Research Laboratory. J.Y.Z and R.F.R acknowledge support from NSF CAREER award CHE:1654732 and generous UCSD startup funds. J.Y.Z acknowledges discussions with Andrea Cavalleri on the various sources of anharmonicities.

***Author contributions.*** B.X. A.D.D. J.W. Y.L. and W.X. designed and performed experiments, B.X. and W.X analyzed experimental data, R.R., A.D.D., B.S.S., J.C.O., J.Y. and W.X. developed the theoretical model, B.X., R.R., A.D.D., B.S.S., J.C.O, J.Y. and W.X. wrote the manuscript.

***Competing financial interests.*** The authors declare no financial conflicts.

***Reference***

1. Muallem, M., Palatnik, A., Nessim, G. D. & Tischler, Y. R. Strong Light-Matter Coupling and Hybridization of Molecular Vibrations in a Low-Loss Infrared Microcavity. *J. Phys. Chem. Lett.* **7,** 2002–2008 (2016).

2. Shalabney, A. *et al.* Enhanced Raman Scattering from Vibro-Polariton Hybrid States. *Angew. Chem. Int. Ed. Engl.* **54,** 7971–7975 (2015).

3. Long, J. P. & Simpkins, B. S. Coherent Coupling between a Molecular Vibration and Fabry – Perot Optical Cavity to Give Hybridized States in the Strong Coupling Limit. *ACS Photonics* **2,** 130 (2015).

4. Simpkins, B. S. *et al.* Spanning Strong to Weak Normal Mode Coupling between Vibrational and Fabry – Pe ́rot Cavity Modes through Tuning of Vibrational Absorption Strength. *ACS Photonics* **2,** 1460 (2015).

5. Saurabh, P. & Mukamel, S. Two-dimensional infrared spectroscopy of vibrational polaritons of molecules in an Two-dimensional infrared spectroscopy of vibrational polaritons of molecules in an optical cavity. *J. Chem. Phys.* **144,** 124115 (2016).

6. Shalabney, A. *et al.* Cohernet coupling of molecular resonators with a microcavity mode. *Nat. Commun.* **6,** 5981 (2015).

7. Takemura, N., Trebaol, S., Wouters, M., Portella-Oberli, M. T. & Deveaud, B. Polaritonic Feshbach resonance. *Nat. Phys.* **10,** 500–504 (2014).

8. Flick, J., Ruggenthaler, M., Appel, H. & Rubio, A. Atoms and molecules in cavities, from weak to strong coupling in quantum-electrodynamics (QED) chemistry. *Proc. Natl. Acad. Sci.* **114,** 3026–3034 (2017).

9. Ebbesen, T. Hybrid Light–Matter States in a Molecular and Material Science Perspective. *Acc. Chem. Res.* **49,** 2403 (2016).

10. Deng, H., Haug, H. & Yamamoto, Y. Exciton-polariton Bose-Einstein condensation. *Rev. Mod. Phys.* **82,** 1489 (2010).

11. Mulliken, R. S. Electronic structures of polyatomic molecules and valence. II. General considerations. *Phys. Rev.* **41,** 49–71 (1932).

12. Sibert, E. L., Tabor, D. P., Kidwell, N. M., Dean, J. C. & Zwier, T. S. Fermi resonance effects in the vibrational spectroscopy of methyl and methoxy groups. *J. Phys. Chem. A* **118,** 11272–11281 (2014).

13. Spano, F. C. & Silva, C. H- and J-Aggregate Behavior in Polymeric Semiconductors. *Annu. Rev. Phys. Chem.* **65,** 477–500 (2014).

14. Amo, A. *et al.* Superfluidity of polaritons in semiconductor microcavities. *Nat. Phys.* **5,** 805–810 (2009).

15. Kéna-Cohen, S. & Forrest, S. R. Room-temperature polariton lasing in an organic single-crystal microcavity. *Nat. Photonics* **4,** 371–375 (2010).

16. Plumhof, J. D., Stöferle, T., Mai, L., Scherf, U. & Mahrt, R. F. Room-temperature Bose – Einstein condensation of cavity exciton – polaritons in a polymer. *Nat. Mater.* **13,** 247 (2014).

17. Daskalakis, K. S., Maier, S. A. & Murray, R. Nonlinear interactions in an organic polariton condensate. *Nat. Mater.* **13,** 271–278 (2014).

18. Deng, H., Weihs, G., Santori, C., Bloch, J. & Yamamoto, Y. Condensation of Semiconductor Microcavity Exciton Polaritons. *Science (80-. ).* **298,** 199–203 (2002).

19. Dunkelberger, A. D., Spann, B. T., Fears, K. P., Simpkins, B. S. & Owrutsky, J. C. Modified relaxation dynamics and coherent energy exchange in coupled vibration-cavity polaritons. *Nat. Commun.* **7,** 13504 (2016).

20. Houdré, R., Stanley, R. P. & Ilegems, M. Vacuum-field Rabi splitting in the presence of inhomogeneous broadening: Resolution of a homogeneous linewidth in an inhomogeneously broadened system. *Phys. Rev. A* **53,** 2711–2715 (1996).

21. Lindberg, M. & Binder, R. Dark states in coherent semiconductor spectroscopy. *Phys. Rev. Lett.* **75,** 1403–1406 (1995).

22. Herrera, F. & Spano, F. C. Dark Vibronic Polaritons and the Spectroscopy of Organic Microcavities. *Phys. Rev. Lett.* **118,** 1–6 (2017).

23. Vasa, P. *et al.* Real-time observation of ultrafast Rabi oscillations between excitons and plasmons in metal nanostructures with J-aggregates. *Nat. Photonics* **7,** 128 (2013).

24. Takemura, N. *et al.* Two-dimensional Fourier transform spectroscopy of exciton-polaritons and their interactions. *Phys. Rev. B - Condens. Matter Mater. Phys.* **92,** 125415 (2015).

25. Cho, M. Coherent two-dimensional optical spectroscopy. *Chemical Reviews* **108,** 1331–1418 (2008).

26. Arrivo, S. M., Dougherty, T. P., Grubbs, W. T. & Heilweil, E. J. Ultrafast infrared spectroscopy of vibrational CO-stretch up-pumping and relaxation dynamics of W(CO)6. *Chem. Phys. Lett.* **235,** 247–254 (1995).

27. Tokmakoff, A., Sauter, B., Kwok, A. S. & Fayer, M. D. Phonon-induced scattering between vibrations and multiphoton vibrational up-pumping in liquid solution. *Chem. Phys. Lett.* **221,** 412–418 (1994).

28. Wen, P., Christmann, G., Baumberg, J. J. & Nelson, K. A. Influence of multi-exciton correlations on nonlinear polariton dynamics in semiconductor microcavities. *New J. Phys.* **15,** (2013).

29. Pino, J. Del, Feist, J. & Garcia-Vidal, F. J. Quantum theory of collective strong coupling of molecular vibrations with a microcavity mode. *New J. Phys.* **17,** 53040 (2015).

30. Takemura, N. *et al.* Coherent and incoherent aspects of polariton dynamics in semiconductor microcavities. *Phys. Rev. B* **94,** 195301 (2016).

31. Scherer, J. J., Paul, J. B., O'Keefe, A. & Saykally, R. J. Cavity ringdown laser absorption spectroscopy: history, development, and application to pulsed molecular beams. *Chem. Rev.* **97,** 25–51 (1997).

32. O'Keefe, A. & Deacon, D. A. G. Cavity ring-down optical spectrometer for absorption measurements using pulsed laser sources. *Rev. Sci. Instrum.* **59,** 2544–2551 (1988).

33. Koch, S. W., Kira, M., Khitrova, G. & Gibbs, H. M. Semiconductor excitons in new light. *Nat. Mater.* **5,** 523–31 (2006).

34. Tassone, F. & Yamamoto, Y. Exciton-exciton scattering dynamics in a semiconductor microcavity and stimulated scattering into polaritons. *Phys. Rev. B* **59,** 10830–10842 (1999).

35. Mukamel, S. & Nagata, Y. Quantum field, interference, and entanglement effects in nonlinear optical spectroscopy. in *Procedia Chemistry* **3,** 132–151 (2011).

36. Jin, G. R. & Liu, W. M. Collapses and revivals of exciton emission in a semiconductor microcavity: Detuning and phase-space filling effects. *Phys. Rev. A - At. Mol. Opt. Phys.* **70,** 13803 (2004).

37. Khalil, M., Demirdöven, N. & Tokmakoff, A. Coherent 2D IR Spectroscopy: Molecular Structure and Dynamics in Solution. *J. Phys. Chem. A* **107,** 5258–5279 (2003).

38. Herzberg, G., William, J. & Spinks, T. *Molecular Spectra and Molecular Structure: Infrared and Raman spectra of polyatomic molecules*. (Van Nostrand, 1939).

39. Middleton, C. T. *et al.* Two-dimensional infrared spectroscopy reveals the complex behaviour of an amyloid fibril inhibitor. *Nat. Chem.* **4,** 355–60 (2012).

40. Rosenfeld, D. E., Gengeliczki, Z., Smith, B. J., Stack, T. D. P. & Fayer, M. D. Structural dynamics of a catalytic monolayer probed by ultrafast 2D IR vibrational echoes. *Science* **334,** 634–9 (2011).

41. Xiong, W. *et al.* Transient 2D IR spectroscopy of charge injection in dye-sensitized nanocrystalline thin films. *J. Am. Chem. Soc.* **131,** 18040–18041 (2009).

42. Ishizaki, A. & Fleming, G. R. Quantum Coherence in Photosynthetic Light Harvesting. *Annu. Rev. Condens. Matter Phys.* **3,** 333–361 (2012).

43. Eaves, J. D. *et al.* Hydrogen bonds in liquid water are broken only fleetingly. *Proc. Natl. Acad. Sci. U. S. A.* **102,** 13019–13022 (2005).

44. Yuen-Zhou, J. *et al.* Coherent exciton dynamics in supramolecular light-harvesting nanotubes revealed by ultrafast quantum process tomography. *ACS Nano* **8,** 5527–5534 (2014).

45. Garrett-Roe, S. & Hamm, P. Purely absorptive three-dimensional infrared spectroscopy. *J. Chem. Phys.* **130,** 164510 (2009).

46. Li, H., Bristow, A. D., Siemens, M. E., Moody, G. & Cundiff, S. T. Unraveling quantum pathways using optical 3D Fourier-transform spectroscopy. *Nat. Commun.* **4,** 1390 (2013).

47. Turner, D. B., Stone, K. W., Gundogdu, K. & Nelson, K. A. Three-dimensional electronic spectroscopy of excitons in GaAs quantum wells. *J. Chem. Phys.* **131,** 144510 (2009).

48. Anna, J. M., Ross, M. R. & Kubarych, K. J. Dissecting Enthalpic and Entropic Barriers to Ultrafast Equilibrium Isomerization of a Flexible Molecule Using 2DIR Chemical Exchange Spectroscopy. *J. Phys. Chem. A* **113,** 6544–6547 (2009).

49. Gandman, A., Mackin, R., Cohn, B., Rubtsov, I. V. & Chuntonov, L. Two-Dimensional Fano Lineshapes in Ultrafast Vibrational Spectroscopy of Thin Molecular Layers on Plasmonic Arrays. *J. Phys. Chem. Lett.* **8,** 3341–3346 (2017).

50. Shim, S.-H. & Zanni, M. T. How to turn your pump-probe instrument into a multidimensional spectrometer: 2D IR and Vis spectroscopies via pulse shaping. *Phys. Chem. Chem. Phys.* **11,** 748–761 (2009).

# Revealing Hidden Vibration Polariton Interactions by 2D IR Spectroscopy

Bo Xiang[2], Raphael F. Ribeiro[1], Adam D. Dunkelberger[3], Jiaxi Wang[1], Yingmin Li[2], Blake. S. Simpkins[3], Jeffrey C. Owrutsky[3], Joel Yuen-Zhou[1], Wei Xiong[1,2]

[1] Department of Chemistry and Biochemistry, University of California, San Diego, La Jolla, CA

[2] Materials Science and Engineering Program, University of California, San Diego, La Jolla, CA

[3] Chemistry Division, Naval Research Laboratory, Washington, District of Columbia

**Supporting Materials**

**S1. 2D IR spectrometer**

**S2. Formation of Polaritons**
   a. **Cavity Modes**
   b. **Detuning Dependence of Cavity Mode Frequencies**
   c. **Strong Coupling between $W(CO)_6$ $T_{1u}$ Vibrational Mode and Cavity Modes**
   d. **Free Induction Decay**
   e. **Spectral Filter Effect**
   f. **2D IR spectra at 105 ps**

**S3. Classical Theoretical Model**

**S4. Quantum-Mechanical Theoretical Model**

**S5. Dependence of 2DIR Features on Cavity Detuning**
   a. **2DIR response versus detuning of cavity**
   b. **Pump-frequency integrated spectral dependence on cavity detuning**
   c. **Influence from dark state response**
   d. **UP-LP and LP-UP Cross Peaks**
   e. **$\omega_{01}$-LP and $\omega_{01}$-UP Cross Peaks**
   f. **LP and UP Diagonal Peaks**

## S1. 2D IR spectrometer

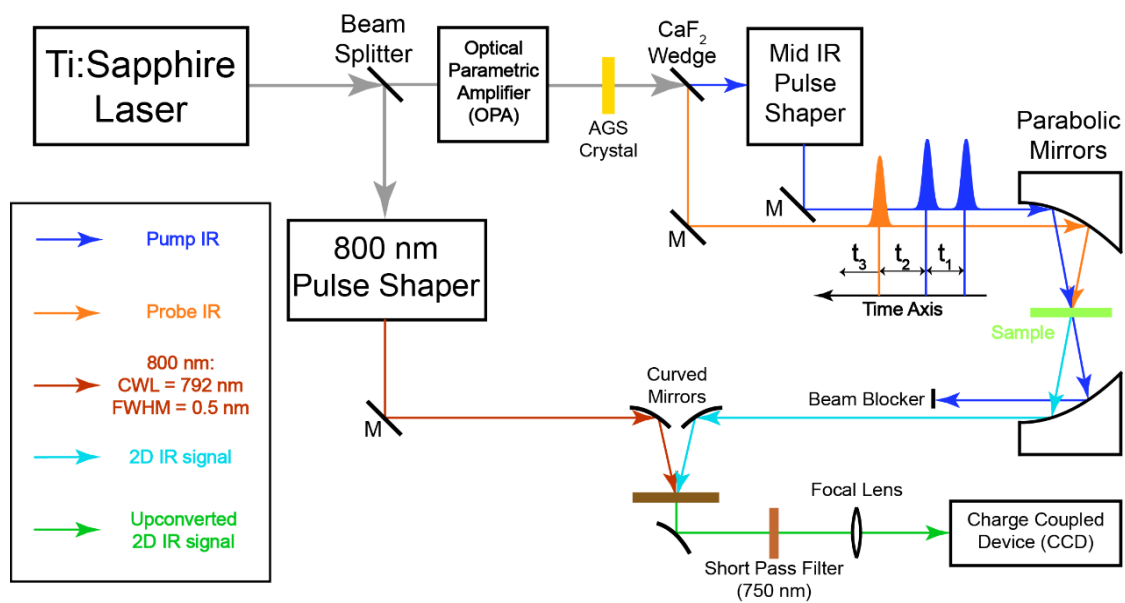

Figure S1. *Scheme of two-dimensional infrared experimental setup.*

Two-dimensional infrared (2D IR) spectroscopy is applied to investigate the light-matter interaction of a $W(CO)_6$/microcavity system. The setup scheme is shown in Figure S1. Initially, 800-nm laser pulses (~35 fs, ~5 W, 1 kHz) are generated by an ultrafast Ti:Sapphire regenerative amplifier (Astrella, Coherent) and sent into an optical parametric amplifier (OPA) (TOPAS, LightConversion) which outputs tunable near-IR pulses. The near-IR pulses are converted to mid-IR pulses through a difference frequency generation (DFG) process by a type II $AgGaS_2$ crystal (Eksma). After DFG, a $CaF_2$ wedge splits the mid IR into two parts: the 95% transmitted part is sent into a Ge-Acoustic Optical Modulator based mid IR pulse shaper (QuickShape, PhaseTech)[1] and is shaped to double pulses, which forms the pump beam arm; the 5% reflected is the probe beam. Both pump (~ 2 µJ) and probe (~ 0.1 µJ) are focused by a parabolic mirror (f = 10 cm) and overlap spatially at the sample. The output signal is collimated by another parabolic mirror (f = 10 cm) at a symmetric position and is upconverted to an 800-nm beam at a 5%Mg: $LiNbO_3$ crystal. The 800-nm beam that comes out of the OPA passes through an 800-nm pulse shaper which narrows its spectrum in the frequency domain (center wavelength of 791 nm and a FWHM of 0.5 nm or 9.5 $cm^{-1}$).

The pulse sequence is shown in Figure S1. Two pump pulses and a probe pulse interact with samples at delayed times ($t_1$, $t_2$ and $t_3$). After the first IR pulse, a vibrational coherence is generated, which is converted into a population state by the second IR pulse and is characterized by scanning $t_1$ (0 to 6000 fs with 20 fs steps) using the mid IR pulse shaper. A rotating frame at $f_0$ = 1583 $cm^{-1}$ is applied to shift the oscillation period to 80 fs and to make the scanning step meet the Nyquist frequency requirement. After waiting for $t_2$, the second coherence is generated by the third IR pulse (probe), which forms a macroscopic polarization that subsequently emits an IR signal. This IR signal is upconverted by a narrow-band 800 nm beam. The upconversion process covers the $t_3$ time delay and the 800-nm pulse duration (full width at half maximum = 0.5 nm) determines the scanning length of $t_3$. The monochromator and CCD (Andor) experimentally Fourier transform the upconverted signal, thus generating a spectrum of the $\omega_3$ axis. Numerical Fourier transform along the $t_1$ axis is required to plot the first coherence along $\omega_1$. The resulting 2D IR spectra are plotted against $\omega_1$ and $\omega_3$. The $t_2$ time delay is scanned by a computerized delay stage which is controlled by home-written LabVIEW programs to characterize the dynamic features of the system. A rotational stage is mounted on the sample stage to control the tilt angle and, therefore, the wavevector space. One special requirement for this experiment is that the rotation axis of the rotational stage needs to be parallel to the incidence plane formed by the pump and probe beams. In this way, we ensure that the in-plane wavevectors, $k_{//}$, of pump and probe pulses are the same. The $k_{//}$ value the of pump and probe beams are determined by checking the 1D transmission polariton spectra of the pump and probe pulses before and after 2D IR acquisitions.

## S2. Formation of Polaritons

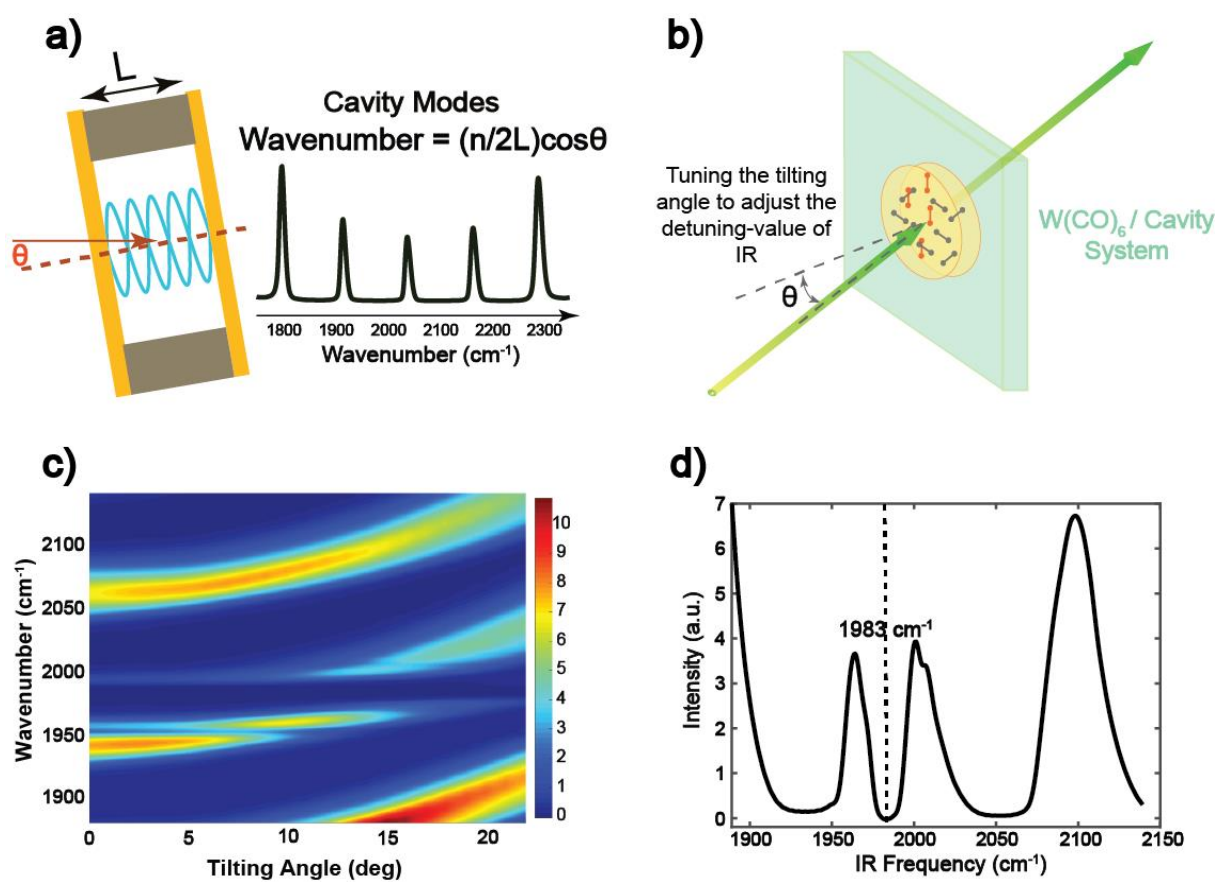

Figure S2. *a) Diagram representing cavity modes; n is the order of mode, L is the cavity length and ϑ is the tilt angle b) Setup scheme of FTIR characterization of molecule-cavity system; c) Dispersion curve of cavity photon-vibration mode coupling; d) FTIR spectrum of polariton under the condition of strongest coupling.*

### (a) Cavity Modes

Cavity modes are standing-wave-like, i.e., only modes of certain frequencies can constructively interfere in the microcavity and form quasi-stationary states. Two dielectric mirrors are used to enclose a microcavity within a spacer of length L, containing a near-saturated $W(CO)_6$ solution. The reflectivity of the mirrors is ~92%, which means the majority of incident photons will be trapped in between the mirrors and bounce back and forth multiple times before leaking to the outside of the cavity. The Q factor of the cavity is ~14. The transmission spectrum is shown in Fig. S2a. The spacing or free spectral range is the same between all neighboring cavity modes; it is $c/2n_cL$ when the incident direction is normal to the sample surface, where c is the speed of light, and $n_c$ is the refractive index of the material in the cavity.

**(b) Detuning Dependence of Cavity Mode Frequencies**

As shown in Fig. S2a, by tuning the tilt angle, the spectrum of the cavity can be controlled to match with the vibrational modes. FTIR characterization (ThermoFischer, Nicolet i10) was used to determine the cavity mode angle dependence. Spectra of cavity modes at different angles were obtained and reflect the variation of the cavity detuning, the energy difference between cavity and the vibration mode ($\Delta = E_{cav} - E_{vib}$) (Fig. S2c). Clearly, as detuning increases, every cavity mode is blue-shifted, and the spacing of neighboring cavity modes shrinks, matching the expected relation in Fig. S2a. The detuning value would have a one-to-one correspondence to the incident angle, θ. This relation can be expressed as[2]

$$E_{LP,UP} = \frac{1}{2}\left\{E_{vib} + E_{cav} \pm \sqrt{4g_0^2 + [E_{vib} - E_{cav}]^2}\right\}. \qquad \text{(Eq. S1)}$$

Hence,

$$E_{UP} + E_{LP} = E_{vib} + E_{cav}.$$

$$\Delta = E_{cav} - E_{vib} = E_{UP} + E_{LP} - 2E_{vib}.$$

**(c) Strong Coupling between $W(CO)_6$ $T_{1u}$ Vibrational Mode and Cavity Modes**

By controlling the incidence angle, the cavity resonance frequency can be adjusted to align with the $T_{1u}$ vibrational frequency (1983 cm$^{-1}$) of $W(CO)_6$ (cavity detuning $\Delta$ = 0). When cavity and vibrational modes are near resonant, and the absorptivity of the vibrational band is high, two polariton branches are formed.[3] The higher-frequency branch is commonly denoted the upper polariton (UP), while the lower frequency modes are lower polaritons (LP). Polariton frequencies are also determined by the incidence angle, i.e., by detuning values. When cavity and molecular vibrational modes are resonant, the intensity of the linear transmission signals for the LP and UP signals are approximately equal, because each type of polariton has nearly equal contributions from photonic and vibrational excitations (Fig. S2d).

**(d) Free Induction Decay**

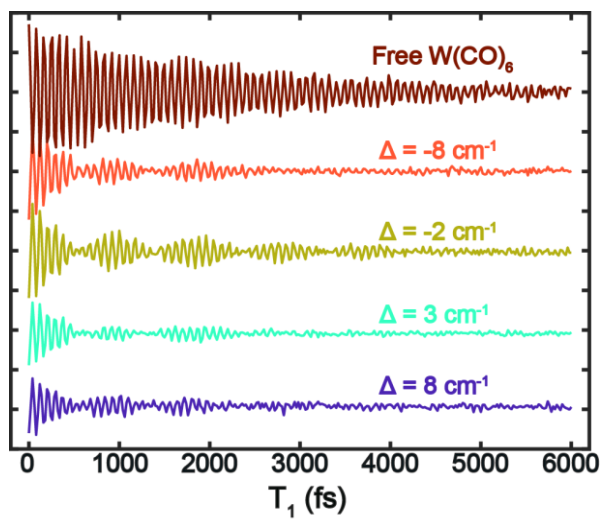

Figure S3. *The free induction decay (FID) of molecules in the microcavity shows beats which vary as a function of cavity detuning.*

The new information from 2D IR can be first learned by inspecting FID of the polariton peaks via a scan of $t_1$ in the time domain. A comparison of FID of the $\omega_{01}$ peak of W(CO)$_6$ in free space and the peak at 1968 cm$^{-1}$ ($\omega_{12}$) in the microcavity (Fig. S3) shows that the dephasing time of vibrational-polaritons is significantly modified relative to decoupled molecular vibrations.

### (e) Spectral Filter Effect

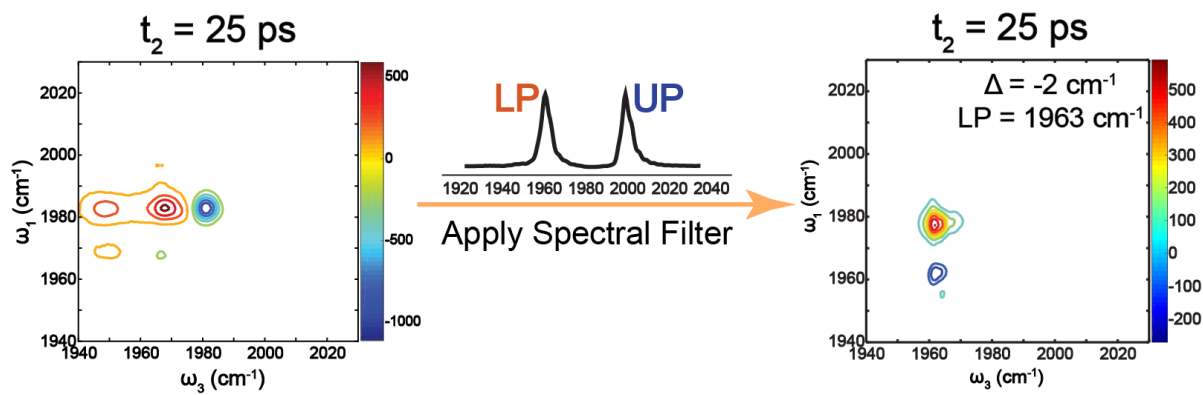

Figure S4. *2D IR spectrum of free molecules ($t_2$ = 25 ps) and its residue after applying a probe polariton (with Δ = -2 cm$^{-1}$) transmission spectral filter.*

2D IR spectra are obtained by performing a Fourier transform FID signals at each frequency pixel along the $t_1$ axis. The 2D IR spectra of polaritons show many peaks due to coupling between different states as confirmed by the observed "quantum beats" in the FID signals (Fig.S3). The third-order emitted signal from the sample is heterodyne-detected by the third pulse, making the latter act as a spectral filter along the $\omega_3$ axis. The spectral signatures which are due exclusively to spectral-filtering can be simulated by utilizing the third pulse as a spectral window on the bare $W(CO)_6$ spectrum, i.e., by convoluting the linear polariton spectrum of the third pulse with the 2D IR spectra of $W(CO)_6$ (Fig. S4, left). The resulting spectrum (Fig. S4, right) is dramatically different from the experimental 2D IR spectra (Fig.3b) except at the regions near $\omega_1$ = 1983 cm$^{-1}$ and $\omega_3$ = 1960 cm$^{-1}$. Thus, the measured signal is unambiguously dominated by the 2D IR vibrational polariton response.

**(f) 2D IR spectra at 105 ps**

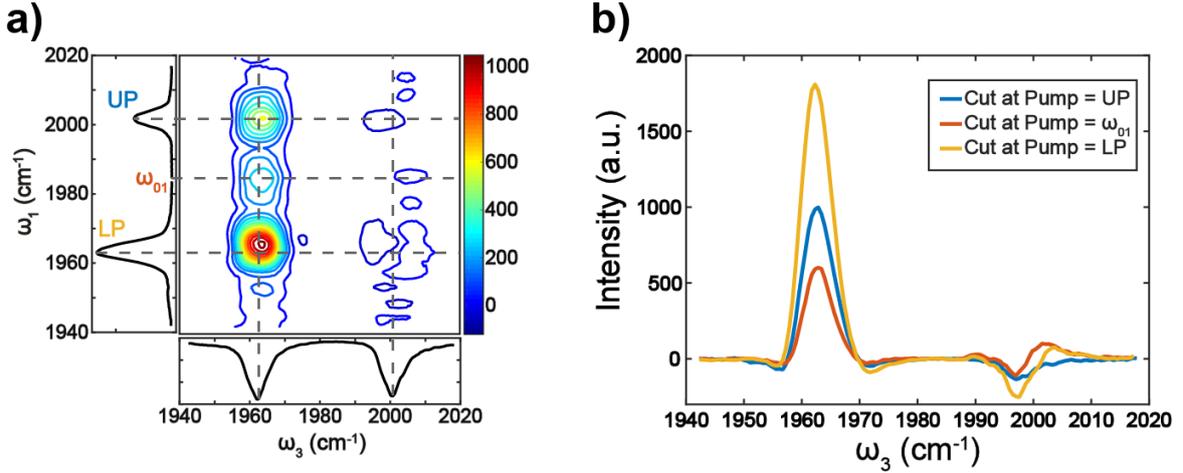

Figure S5 *(a) 2D IR spectrum of polaritons with $\Delta$ = -2 cm$^{-1}$ ($t_2$ = 105 ps); (b) Spectral cuts at UP, $\omega_{01}$, and LP pump frequencies. The position of the 2D IR peaks at both probe LP and UP frequencies are independent of the pump frequency.*

## S3. Classical Theoretical Model

Equation S2, below, is the classical equation for the transmission of a Fabry-Pérot cavity. This expression can provide a basis for relating transient spectra to excited and ground state populations[4].

$$T_{cav}(\bar{\nu}) = \frac{T^2 e^{-\alpha L}}{1+R^2 e^{-2\alpha L}-2Re^{-\alpha L}\cos(4\pi nL\bar{\nu}+2\varphi)} \quad (S2)$$

This relationship is based on the frequency dependent absorption coefficient ($\alpha$) and refractive index (*n*) of the material within the cavity. We obtain $\alpha$ and $n$ by modeling the dielectric function of the cavity load as a sum of Lorentzian oscillators. The real and imaginary components of the dielectric function, $\varepsilon_1$ and $\varepsilon_2$, are defined as a sum of *i* Lorentzian oscillators according to

$$\varepsilon_1 = n_{bg}^2 + \sum_i \frac{A_i(\nu_{0i}^2-\nu^2)}{(\nu_{0i}^2-\nu^2)^2+(\Gamma_i\nu)^2}, \quad \text{and} \quad (S3)$$

$$\varepsilon_2 = \sum_i \frac{A_i \Gamma_i \nu}{(\nu_{0i}^2-\nu^2)^2+(\Gamma_i\nu)^2}, \quad (S4)$$

where $n_{bg}$ is the background refractive index, $A_i$ the amplitude, $\nu_{0i}$ the resonant frequency,

and $\Gamma_i$ the full linewidth associated with the $i^{th}$ oscillator. The frequency-dependent refractive index, $n$, and absorption coefficient, $\alpha$, can be formulated as

$$n = \sqrt{\frac{\varepsilon_1 + \sqrt{\varepsilon_1^2 + \varepsilon_2^2}}{2}}, \quad (S5)$$

$$\alpha = 4\pi\nu k = 4\pi\nu\sqrt{\frac{-\varepsilon_1 + \sqrt{\varepsilon_1^2 + \varepsilon_2^2}}{2}}. \quad (S6)$$

Initial values of $A_i$, $\nu_{0i}$, and $\Gamma_i$ are chosen to be consistent with the optical response of witness samples, *i.e.*, absorbance for the concentration and pathlengths used. Expressions S3 and S4 are then substituted into S5 and S6 which are substituted into S2 giving the transmission spectrum through a Fabry-Pérot cavity with tailorable parameters describing the oscillator resonance frequencies, linewidths, and amplitudes. We use 96% for the reflectivity of the mirrors and 4% for the transmission. Transmission spectra may be generated for a system with a fraction of the ground state population excited to the first excited state by reducing $A_0$ and increasing $A_1$ in expression S3 and S4 above. Comparing this excited-state spectrum with the spectrum associated with an entirely ground-state system gives a prediction for the transient spectrum. The example data below (Fig. S6) shows the ground state spectrum (blue), excited state spectrum (5% of ground state population excited to $\nu_1$; red), and the transient spectrum (defined as $-(T_{excited}-T_{ground})$; scaled for visibility and shown in magenta).

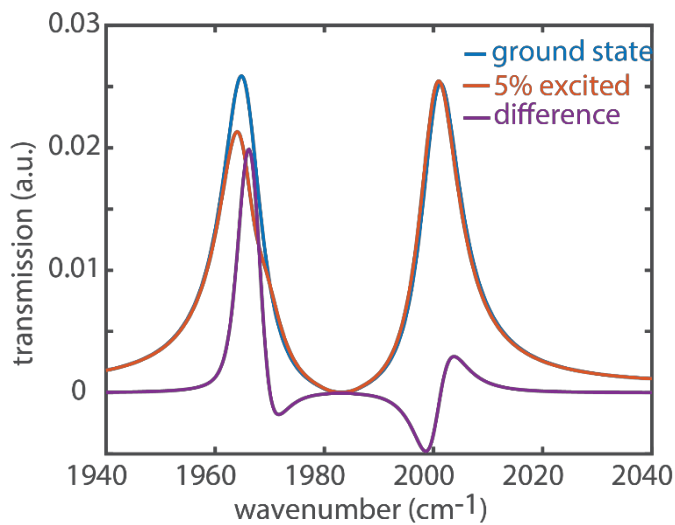

Figure S6. *Polariton transmission spectra when the system is in ground state and 5% of excited states is populated, and the difference spectra.*

Eq. S2 represents the special case (single layer) of the general transfer matrix approach for calculating transmission and reflection through a stack of layers. This approximation works well for predicting transmission spectra but falters for quantitative reflection. We will now apply the complete transfer matrix approach to our multilayered system to calculate transmission and reflection, (and thus absorption since A=1-T-R) for the cavity coupled system. We use the dielectric function of $W(CO)_6$, generated through Eqs. S3 and S4, and the dielectric properties of the layers comprising the dielectric stack mirrors (these are proprietary but we assume Ge and $Si_3N_4$). We find two notable results with implications for the current work. (Fig.

S7) First, there is notable absorption at the fundamental ν$_{01}$ frequency located directly between the upper and lower polaritons. This absorptive feature is likely related to direct excitation of reservoir states as discussed in the main text (see Fig. 3 and discussion). Secondly, if we define the LP and UP frequencies as the peaks found in the transmission spectra, we can extract the polariton absorption (defined as the 1-T-R at the LP and UP positions) as a function of cavity tuning, θ. These absorption values, along with the polariton separation, are plotted below. One sees that polariton absorption is maximized very near the condition of zero detuning (*i.e.*, angle at which polariton separation is minimized, θ~18°). This behavior substantiates our proposition that polaritons are most efficiently excited when cavity detuning is minimized (see discussion of Fig. 4b and c in main text).

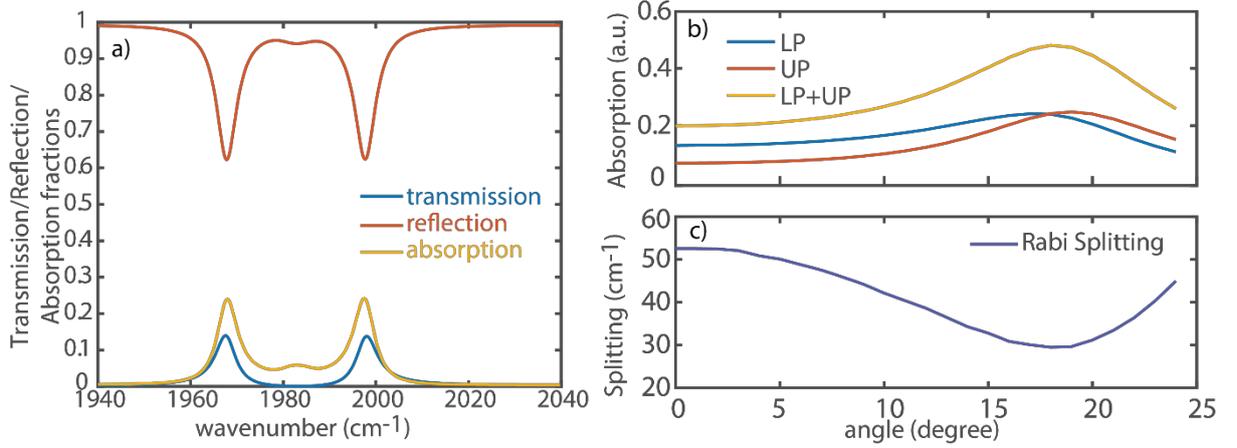

Figure S7. *(a) Modeled transmission, reflection and absorption spectra at zero detuning, and (b) absorption coefficient and (c) Rabi splitting as a function of tilt angle.*

## S4. Quantum-Mechanical Theoretical Model

Our minimal model for the W(CO)$_6$-cavity system includes the free Hamiltonian of *N* independent anharmonic vibrations, a single cavity-mode, the strong light-matter interaction, and the weak coupling of the external electromagnetic field modes with the cavity photon. The interaction of the probed vibrations with the environment is accounted for by a phenomenological damping constant obtained from the linewidth of the bare molecule 0-1 band. Thus, the total Hamiltonian of the system can be expressed as

$$H = H_{\mathrm{mol}} + H_{\mathrm{cav}} + H_{\mathrm{mol\text{-}cav}} + H_{\mathrm{cav\text{-}ext}},  \quad \text{(Eq.S7)}$$

where $H_{\mathrm{mol}} + H_{\mathrm{cav}}$ contains the bare-molecule and empty cavity dynamics,

$$H_{\mathrm{mol}} + H_{\mathrm{cav}} = \sum_{i=1}^{N} \left( \hbar\omega_0 a_i^\dagger a_i - \hbar\Delta_{\mathrm{anh}} a_i^\dagger a_i^\dagger a_i a_i \right) + \hbar\omega_c b^\dagger b, \quad \text{(Eq.S8)}$$

where $\omega_0 (\omega_c)$ is the molecular (cavity) fundamental transition frequency (1983 cm$^{-1}$), $\Delta_{\mathrm{anh}}$ gives rise to the energetic detuning of the excited-state relative to the ground state, i.e., $\hbar\omega_{12} = \hbar\omega_0 - 2\hbar\Delta_{\mathrm{anh}} = 1968$ cm$^{-1}$, and $a_i$ and $b$ are the *i*th molecular and cavity-mode annihilation operators, respectively. The interaction between cavity and molecular degrees of freedom is described by

$$H_{\text{mol-cav}} = \hbar g \sum_{i=1}^{N} \left( b^{\dagger} a_i + a_i^{\dagger} b \right) + \hbar g \delta \sum_{i=1}^{N} \left( b^{\dagger} a_i^{\dagger} a_i a_i + a_i^{\dagger} a_i^{\dagger} a_i b \right).$$

(Eq.S9)

$H_{\text{mol-cav}}$ includes the strong coupling between molecular vibrations and cavity modes: the first term is the typical rotating wave approximation (RWA) approximation to the cavity-matter dipolar interaction with frequency $g$, while the second contains the anharmonic corrections to this interaction due to non-Condon effects[5], i.e., $\delta$ is the electrical anharmonicity parameter, and it can be related to the 1-2 transition dipole moment via $\mu_{12} = \sqrt{2}\mu_{01}(1 + \delta)$[7]. The coupling of cavity and external electromagnetic modes is assumed weak and frequency-independent, so it can be modeled according to Markovian input-output theory[5]. The last term in Eq. S7, $H_{\text{cav-ext}}$ represents the weak (linear) interaction between the cavity modes and the continuous spectrum of electromagnetic (EM) modes outside of the cavity, which enables excitation and detection in 2D spectroscopy and ensures that theories for 2D spectroscopy based on weak perturbations are still suitable for polariton systems.

The linear transmission spectrum of the cavity-molecule system can be obtained by solving the Heisenberg equations of motion for the cavity modes in terms of the (classical) input laser pulse (described by $b_{\text{in}}^{L}$) impinging on the cavity left mirror, and then relating the right-hand-side output modes ($b_{\text{out}}^{R}$) to the internal cavity field. The output field spectrum, in terms of the incoming can be written as

$$b_{\text{out}}^{R}(\omega') = \frac{\kappa/2[i(\omega' - \omega_0) - \gamma_m/2]}{[i(\omega' - \omega_c)][i(\omega' - \omega_0) - \gamma_m/2] + Ng^2} b_{\text{in}}^{L}(\omega'),$$

(Eq.S10)

where $\gamma_m$ is the FWHM of the fundamental molecular transition, and $\kappa/2$ is the width of the cavity as a consequence of the cavity coupling to the left and right external EM modes. The transmission spectrum is directly proportional to the output electric field power spectrum $|b_{\text{out}}^{R}(\omega')|^2$.

A detailed model of the nonlinear response of the system requires the inclusion of its complete dynamics under the driving exerted by the pump and probe pulses. Given the many-body interactions which emerge between polaritons and dark-states when the Hamiltonian is written in the delocalized representation[6,7], we pursue here an alternative approach. Noting that the lifetime of the cavity mode is less than 5 ps, we suppose that at a probe delay time of 25 ps, the average number of populated cavity photons is zero, but a transient excited-state population exists. This is equivalent to supposing that only reservoir/dark-modes are appreciably occupied when $t_2 > 25$ ps. Additionally, we also suppose that any transient coherence can be neglected, and no v > 1 states are occupied. Thus, on the assumption the probe pulse is sufficiently weak, we obtain an approximation for the pump-probe transmission spectrum by solving the Heisenberg-Langevin input-output equations with the initial condition described above (for the detailed mathematical treatment including analytical expressions for the pump-probe transmission spectrum, see Ref. [8]). Below we provide the results of numerical simulations of the differential pump-probe transmission (Fig. S8) and probe transmission spectra (Figure S8) for three different detuning values of $\Delta = \omega_c - \omega_0$, assuming the average (pump-induced) transient population of excited-state molecules at $t_2$ is $N_{\text{pump}} = \sum_{i=1}^{N} N_i(t_2) = 0.05N$, where $N$ is the number of molecules in the cavity, the electrical anhamonicity parameter is $\delta = -0.25^9$, the cavity linewidth $\kappa = 10$ cm$^{-1}$ and the vibrational fundamental transition linewidth $\gamma_m = 3$ cm$^{-1}$.

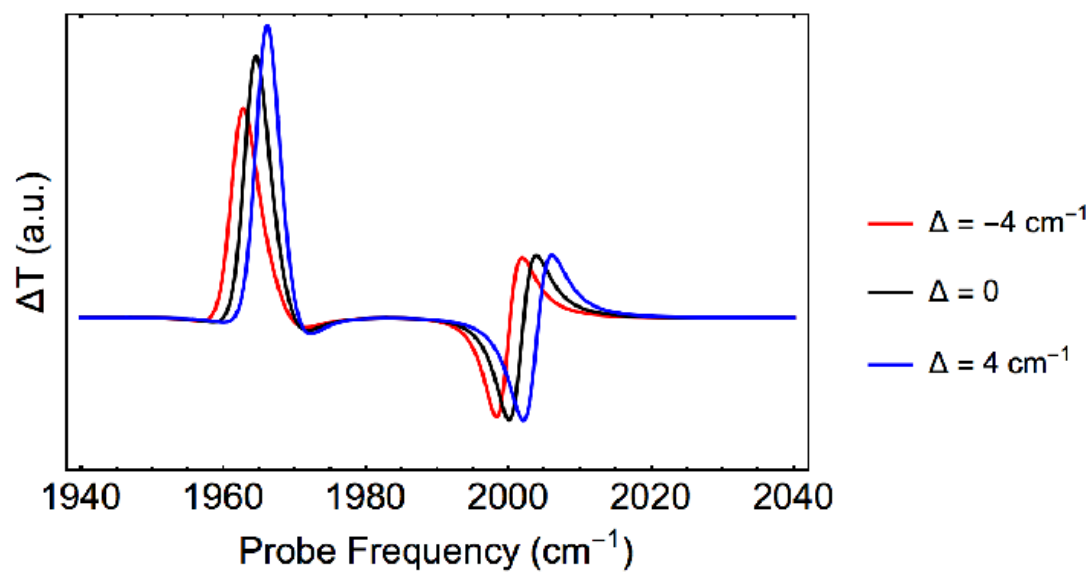

Figure S8. *Differential probe linear transmission spectra for different cavity-vibration detunings with pure molecular transient population* $N_{\text{pump}} = 0.05N$

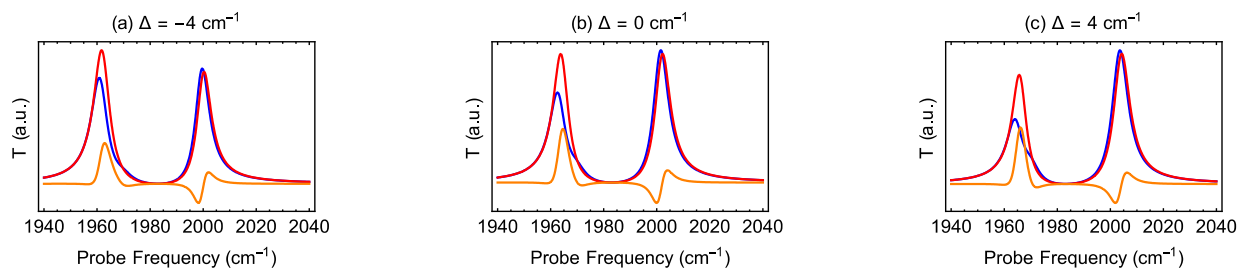

Figure S9. Linear probe transmission in the presence (blue) and absence (red) of a 5% transient population of reservoir molecular excitation, and the corresponding difference transmission spectra (orange).

Figs. S8 and S9 display behavior which agrees with the experimental pump-probe spectrum of Fig. 2b of the main text. The simulations are also consistent with the main trends of Fig. 4a, i.e., the theoretical spectra explain the red-shift of the UP peaks as well as the reduction in the intensity of the large ($\Delta T$) signal near the LP as the cavity is red-tuned. It can be seen from Fig. S8 that as the LP is tuned to be near resonant with $\omega_{12}$, (i.e., dark reservoir transitions) the impact of the transient population is maximized. This can be seen by comparing the red and blue curves in Fig. S9, where the difference is most striking when the LP frequency is tuned to be near-resonant with $\omega_{12}$ (see Fig. S9c). The theory here described reveals the UP signals are much less sensitive to the molecular transient population than the LP. This can be rationalized by the fact that nuclear anharmonicity makes the bare molecule 1-2 transition red-shifted from the fundamental, so it is far off-resonant with the UP. Nonetheless, a clear red-shift of the UP can be observed in Figure S9, in agreement with the experimental results reported in this paper.

We also note that the three resonances of the transient transmission spectrum (blue curves in Fig. S9) can be represented as eigenvalues of a 3x3 mode-coupling matrix, where the rows/columns correspond to oscillators with frequencies $\omega_c, \omega_{01}$, and $\omega_{12}$. If for simplicity we neglect the bare molecule linewidth, then the mode-coupling matrix containing the transient resonances predicted by the QM model is given by

$$h_{\text{eff}}(f^{\text{pu}}) = \begin{pmatrix} \omega_c - i\kappa/2 & g\sqrt{N}\sqrt{1-2f^{\text{pu}}} & g\sqrt{2f^{\text{pu}}N}(1+\delta) \\ g\sqrt{N}\sqrt{1-2f^{\text{pu}}} & \omega_0 & 0 \\ g\sqrt{2f^{\text{pu}}N}(1+\delta) & 0 & \omega_0 - 2\Delta_{\text{anh}} \end{pmatrix},$$ (Eq. S11)

where $f^{\text{pu}} = N_{\text{pu}}/N$. The above matrix may be interpreted to arise from the interaction of the cavity photon with classical matter polarization resulting from 0-1 and 1-2 vibrational transitions. Thus, the diagonal elements contain the bare resonances, and the off-diagonal couplings correspond to the cavity coupling with the two sources of molecular polarization. Note that if $N_{\text{pu}} = 0$, only the 0-1 mode is coupled to the cavity, and we recover the linear response Rabi splitting. The 2x2 block on the l.h.s illustrates the Rabi splitting contraction due to a reduction in the ground-state population. The $[h_{\text{eff}}(f^{\text{pu}})]_{13}$ and $[h_{\text{eff}}(f^{\text{pu}})]_{31}$ elements contain the interaction of reservoir excited-state absorption with the cavity photon. These elements are responsible for the largely asymmetric nonlinear response of vibration-polaritons even when the cavity-vibrational detuning vanishes (if $[h_{\text{eff}}(f^{\text{pu}})]_{13}$ vanished, the 1-2 molecular vibration would be decoupled from the cavity, and there would only be Rabi splitting contraction as it happens with many inorganic semiconductor exciton-polaritons when Pauli blocking or phase-space filing is the only significant source of polariton anharmonicity).

Finally, the interpretation of the QM model in terms of a classical mode-coupling matrix explains the efficacy of the classical description given in Section 3: in fact, the QM and classical models become similar in the limit studied here, as in both cases the transient transmission spectra can be attributed to the coupling of cavity modes with matter polarization arising from ground-stable bleach and stimulated emission of the 0-1 transition and excited-state absorption of the 1-2 transition. In the figure below we display Feynman diagrams for the matter polarization which interacts with the cavity mode and lead to the resonances observed in the transient spectrum.

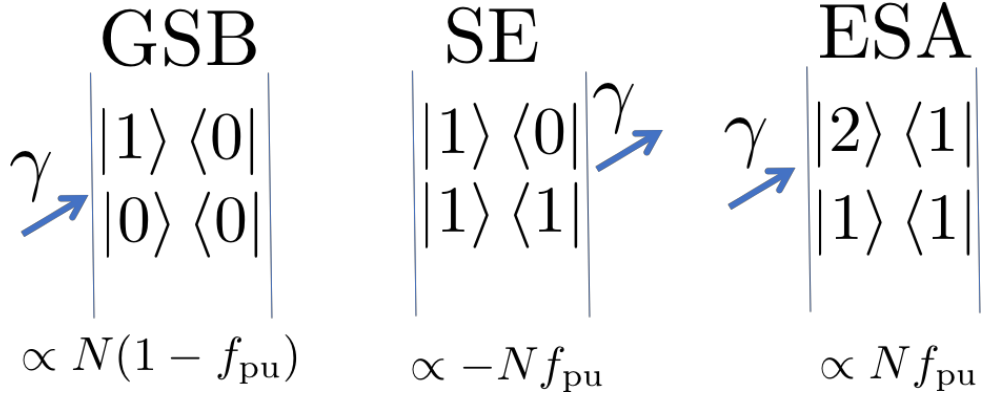

Figure S10. *Feynman diagrams for ground-state bleach, stimulated emission and excited-state absorption contributions to the matter polarization. These interact with the cavity according to the couplings indicated in Eq. S11 and lead to the resonances observed in the pump-probe transmission spectra in Figs. S8-S9.*

The exact energy spectrum of the model utilized in this work is complicated due to the self and cross polariton-polariton and dark-state interactions induced by electrical and mechanical anharmonicities. Nonetheless, the nonlinear spectra presented above reveal a simple *effective* picture for the nonlinear response. Particularly, we may utilize the resonances of the theoretical response function to draw a set of effective energy levels (Fig.S11), where each of them corresponds to a resonance in the excited-state spectrum. As explained above, these peaks, and much of the response predicted with this QM model. Below we show the energy levels which correspond to the center of the bands obtained by the quantum-mechanical numerical simulation and reported in Figs. S8 and S9. We note this QM picture is only an approximated picture in aim to bridge the complete quantum picture to the established response function theory.

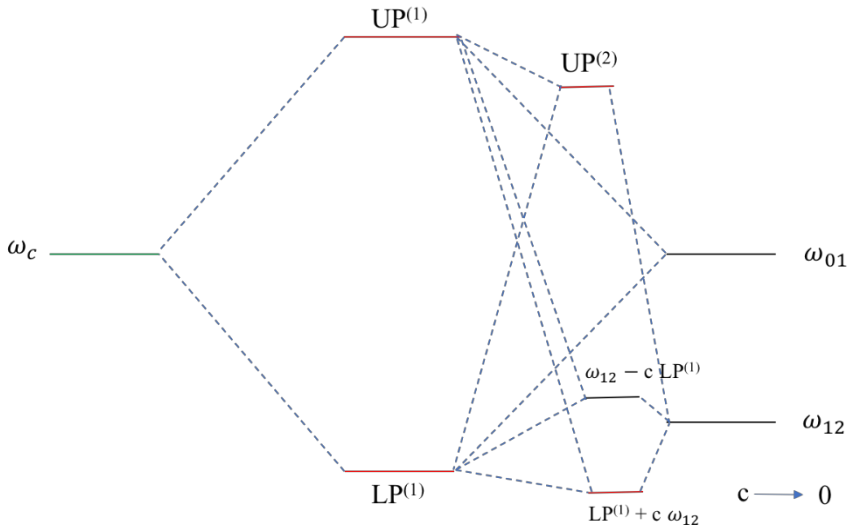

Figure S11. *Effective energy diagram for the QM model.*

## S5. Dependence of 2DIR Features on Cavity Detuning

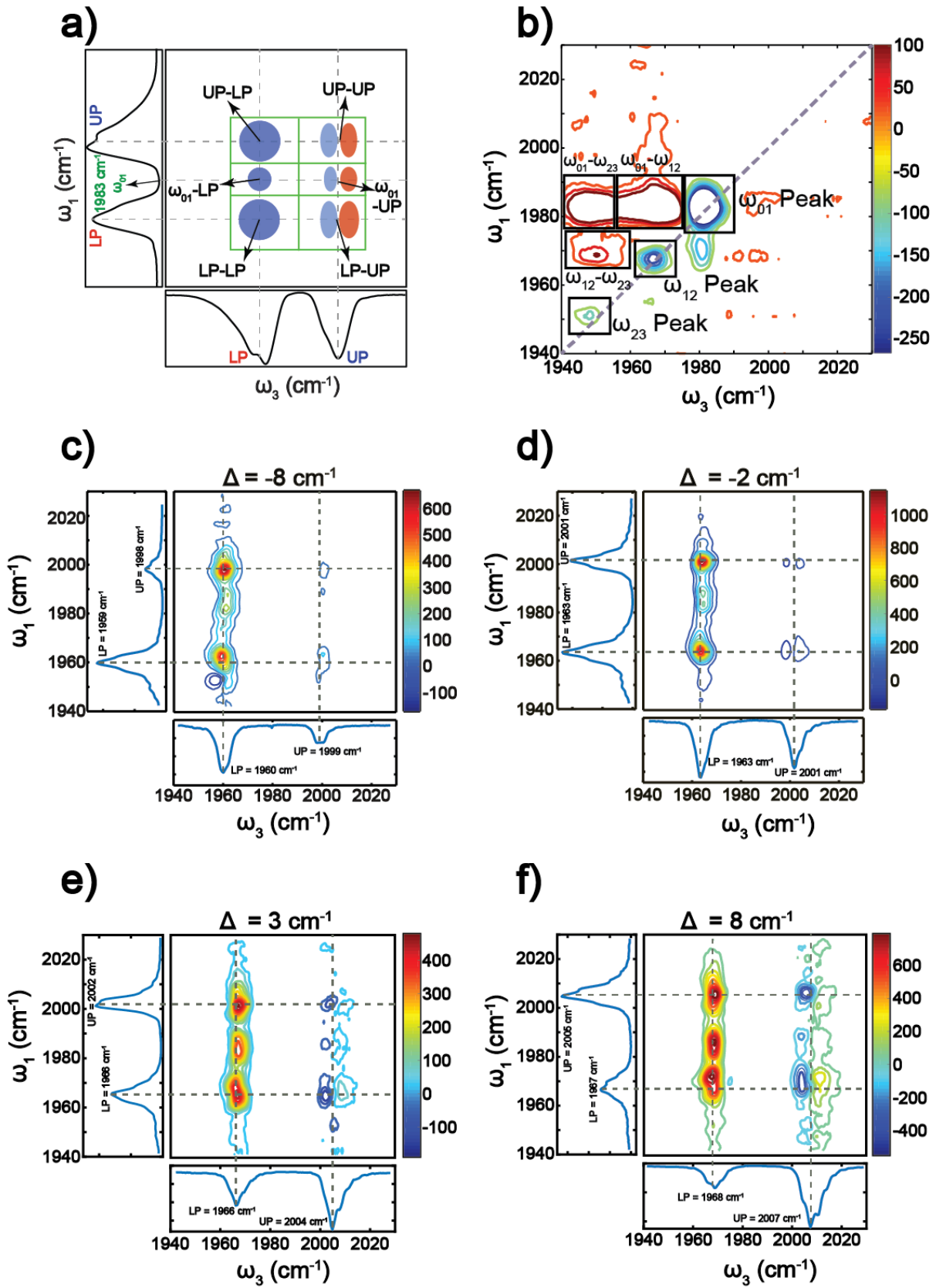

Figure S12. *2DIR spectra ($t_2$ = 25 ps) of (a) generic polaritons; (b) Free $W(CO)_6$ molecules; Polaritons with (c) Δ = -8 $cm^{-1}$; (d) Δ = -2 $cm^{-1}$; (e) Δ = 3 $cm^{-1}$; (f) Δ = 8 $cm^{-1}$.*

### (a) 2DIR response versus detuning of cavity

Figure S12a shows the typical six regions in polariton 2DIR spectra which can be identified based on the $\omega_1$ (pump IR) and $\omega_3$ (probe IR) coordinates of corresponding peaks. These regions include UP-LP, dark mode-LP ($\omega_{01}$-LP), LP-LP, UP-UP, dark mode-UP ($\omega_{01}$-UP) and LP-UP. The positions of all six peaks have different cavity detuning dependence, based on their origin, and deviations from the expected dependence also occur (detailed analysis is provided in later sections). By analyzing the positions of the peaks and their intensities as a function of detuning (Fig. 4 in the main manuscript), the physical origin of the six major peaks can be attributed.

### (b) Pump-frequency integrated spectral dependence on cavity detuning

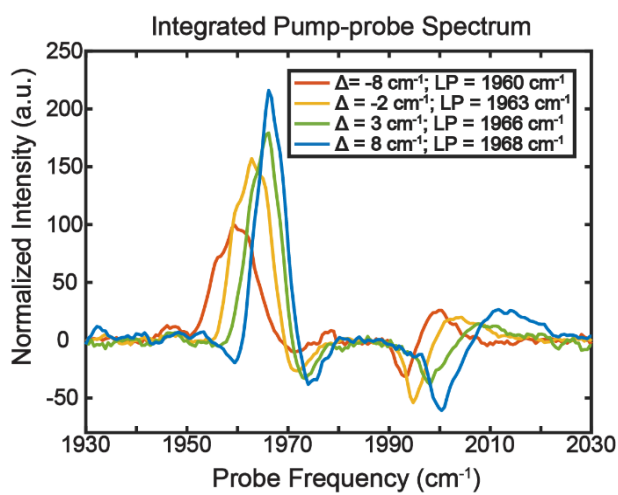

Figure S13. *Integrated pump-probe spectrum for different detuning values. The peaks contain contributions of probe transitions involving UP, dark modes and LP.*

### (c) Influence from dark state response

There are dark states that can be weakly excited by infrared excitation that is not evident in 1D pump - probe studies but is unambiguously demonstrated in the 2DIR measurements. To determine the frequencies of these pure dark state peaks, a 2DIR experiment was done on a bare $W(CO)_6$ hexane solution (~7 mM), i.e., outside of a microcavity, as a control experiment (Fig. S12b). Besides the fundamental and excited state resonances, high-order peaks appear. To determine the dark state residue, a polariton spectrum was applied as spectral window (Fig. S4). These results indicate that dark state response may also appear in the vibrational-polariton spectra, in the LP-LP, $\omega_{01}$-LP and UP-LP regions. Based on the polariton spectral window, the following peaks labeled in Fig. S9b could appear in the 2D spectra: $\omega_{12}$ and $\omega_{23}$ diagonal peaks and $\omega_{01}$-$\omega_{12}$, $\omega_{01}$-$\omega_{23}$ and $\omega_{12}$-$\omega_{23}$ cross peaks. In practice, these dark mode responses can be somewhat suppressed by carefully detuning the cavity mode to reduce spectral overlap between polaritons and bare molecule resonances. For instance, with a suitable LP frequency (e.g., when LP ~1962 cm$^{-1}$, the overlap with either $\omega_{12}$ (1968 cm$^{-1}$) or $\omega_{23}$ (1951 cm$^{-1}$) is reduced), the pure molecule response is mitigated, although the broad spectral width of $\omega_{12}$ still implies it contributes to the spectral features in the LP-probe region. In the following sections, cavity detuning dependence is analyzed and the results provide convincing evidence on the character of the spectral peaks.

To determine whether the observed 2D peaks originate from purely polariton excitations, polariton-dark mode interactions, or pure dark state excitation, we examined the dependence of the observed resonances on the cavity detuning. The variation of the position of the peaks is an effective way to examine the nature of the observed resonances, because in general, polariton peaks shift as the cavity is detuned, whereas dark state resonances remain fixed. This is well demonstrated by measurements of FTIR and polariton dispersion curves (Fig. S2c). Similarly, if the peaks of the 2D spectra arise from the excitation of polaritons, the projection of their positions onto the pump and probe axes should shift with detuning, in the same way peaks shift with respect to detuning in the 1D polariton transmission spectra. On the other hand, if the 2D spectra arise purely from dark states, the projection of its peaks on the pump axis should be insensitive to detuning. The above arguments on the dependence of polariton frequencies with respect to detuning hold very well for projections on the pump axis, and also work for peaks located around the UP along the probe frequency. For the resonances located near LP on the probe axis, the detuning dependence is complicated, because spectral window effects can play a role by enabling the observation of additional hot band transitions of reservoir modes. Yet, because we can unambiguously determine the physical origin of spectral peaks appearing near the probe UP frequency, and the spectral shifts for LP and UP at the probe axis are directly correlated, we can infer that the contribution of pure LP transition at spectral peaks appears around the probe LP frequency.

We will show that in our spectra, both cases (peaks involving pure polariton transitions, and pure dark states), as well as some intermediate scenarios coexist. We choose to plot the correlation between the positions of the peaks in the 1D polariton transmission spectrum, and the projections of the resonances observed in the 2D spectra (Fig.S14-S16). In this way, we expect the 2D spectra polariton resonances to shift similarly to the 1D spectrum polariton peaks, forming a linear relationship.

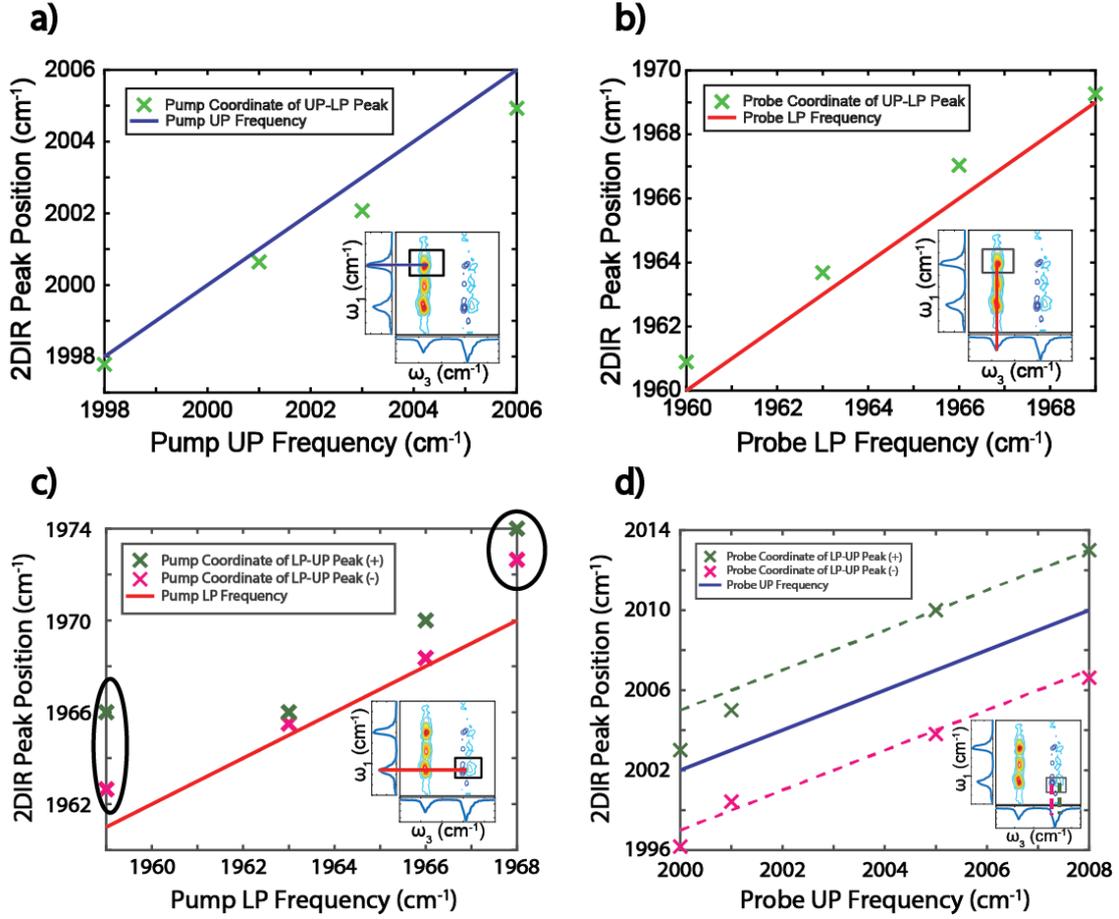

Figure S14 (a) UP-LP peak: 2D spectral peak position projection onto $\omega_1$ axis versus UP peak frequency in 1D pump transmission spectrum; (b) UP-LP peak: 2D spectral peak position projection onto $\omega_3$ axis versus LP peak frequency in 1D probe transmission spectrum; (c) LP-UP peak: 2D spectral peak position projection onto $\omega_1$ axis versus LP peak frequency in 1D pump transmission spectrum; (d) LP-UP peak: 2D spectral peak position projection onto $\omega_3$ axis versus UP peak frequency in 1D probe transmission spectrum.

### (d) UP-LP and LP-UP Cross Peaks

We first analyze the UP-LP and LP-UP cross peaks. When UP and LP are excited, they interact with each other either directly[10] or through populating dark states, and thus it is expected that UP-LP and LP-UP cross peaks form. However, the UP-LP cross peaks may also be influenced by dark-mode $\omega_{12}$ response. Based on Figure S14a and S14b, the UP-LP peak position on both pump and probe axes match well with the corresponding polariton peak position measured in the 1D transmission spectra of pump and probe beams (form linear relationship). Thus, we can conclude that UP-LP peak are excited through purely UP states. The optical transitions that give rise to the peaks at the LP frequency along probe axis are more complicated, because LP and $\omega_{12}$ are close in frequency, and thus these modes interact to give the observed response (see Fig. S11). Nonetheless, as further described in (f) (on the physical origin of UP-UP peaks) it is clear that excitation of the UP changes its optical response, and given the correlated nature of UP and LP, this implies the LP is also affected when the UP mode is excited. Thus, we infer that when exciting UP, the excitation must also alter the optical response of LP. The theoretical models presented in Secs. 3 and 4 support this notion.

The LP-UP cross peaks show similar linear relationship with the 1D polariton transmission spectral position, except in the case of Figure S14c, where deviations towards higher frequencies are observed when the LP is maximally red- or blue-detuned (black circles in figure S14c). We speculate that when polariton modes are closer to dark states, the latter appear through weak coupling, and higher-order peaks from the reservoir molecules (see Fig. S12b) complicate the 2DIR response. These peaks which have different origins merge into each other, causing the shift of the major peak (in face of the overlap of multiple peaks). In the case of LP-UP cross peaks, possibly, $\omega_{12}$-UP and $\omega_{23}$-UP cross peaks interfere with the LP-UP cross peaks and cause the observed blue-shift.

In general, the LP-UP peak originates mainly from pumping and probing polariton transitions, and the UP-LP peak has a mixed contribution from pumping polariton transitions and probing the dark states, and pumping and probing polaritons. We note dark states cease to be "dark", because they hybridize with the polariton modes and therefore carry some polaritonic features as well (see Sec. 4).

**(e) $\omega_{01}$-LP and $\omega_{01}$-UP Cross Peaks**

These two peaks are important because they unambiguously demonstrate excitation of reservoir states which is not evident in the 1D spectra.[4] The decoupled nature of the dark ($\omega_{01}$) states is shown not only by their pump axis frequency (as shown below) but also by the fact that the corresponding resonances are not dispersive; unlike the UP and LP, the $\omega_{01}$ transitions are independent of the incidence angle. These features also indicate interaction between dark-modes and polaritons.[11–13] In Fig. S15a, the projections of $\omega_{01}$-LP peaks onto pump axis remain constant at the $\omega_{01}$ frequency, in sharp contrast with the shift of LP frequency from 1957 cm$^{-1}$ to 1967 cm$^{-1}$ in the 1D pump beam spectrum, which indicates that the coherence involved in the pump $t_1$ period is associated to dark mode transitions. On the other hand, the projection of the $\omega_{01}$-LP peak onto the probe axis (Fig. S15b) shifts with respect to the LP frequency measured in the 1D probe spectrum, just as predicted above. However, the peak intensity- detuning relationship (Fig. 5a) indicates a different scenario. The intensity of the $\omega_{01}$-LP peak is enhanced when the probe-LP frequency approaches $\omega_{12}$ at 1968 cm$^{-1}$. This trend implies the coherence detected by the probe beam has a sufficiently strong dependence on $\omega_{12}$ transitions, because as the LP mode moves closer to $\omega_{12}$, the mixing of the latter with the former is accentuated and thus the peak becomes more intense. Thus, both the pump-frequency dependence and the intensity trend indicate that the $\omega_{01}$-LP peak is strongly affected by dark state response (such as residue of $\omega_{01}$- $\omega_{12}$ in Fig. S11b). With such strong contributions, the precise influence of pumping $\omega_{01}$ on the LP transition is hard to be decoded experimentally. However, as shown below, dark state excitation can influence UP transitions, and, since there are strong coherent interactions between UP and LP, we believe pumping the dark state should also give rise to a shifted LP transition. The theoretical model presented in Sec. 4 is consistent with these notions, as it reveals that sufficiently large dark-mode population will have a significant effect on LP and UP.

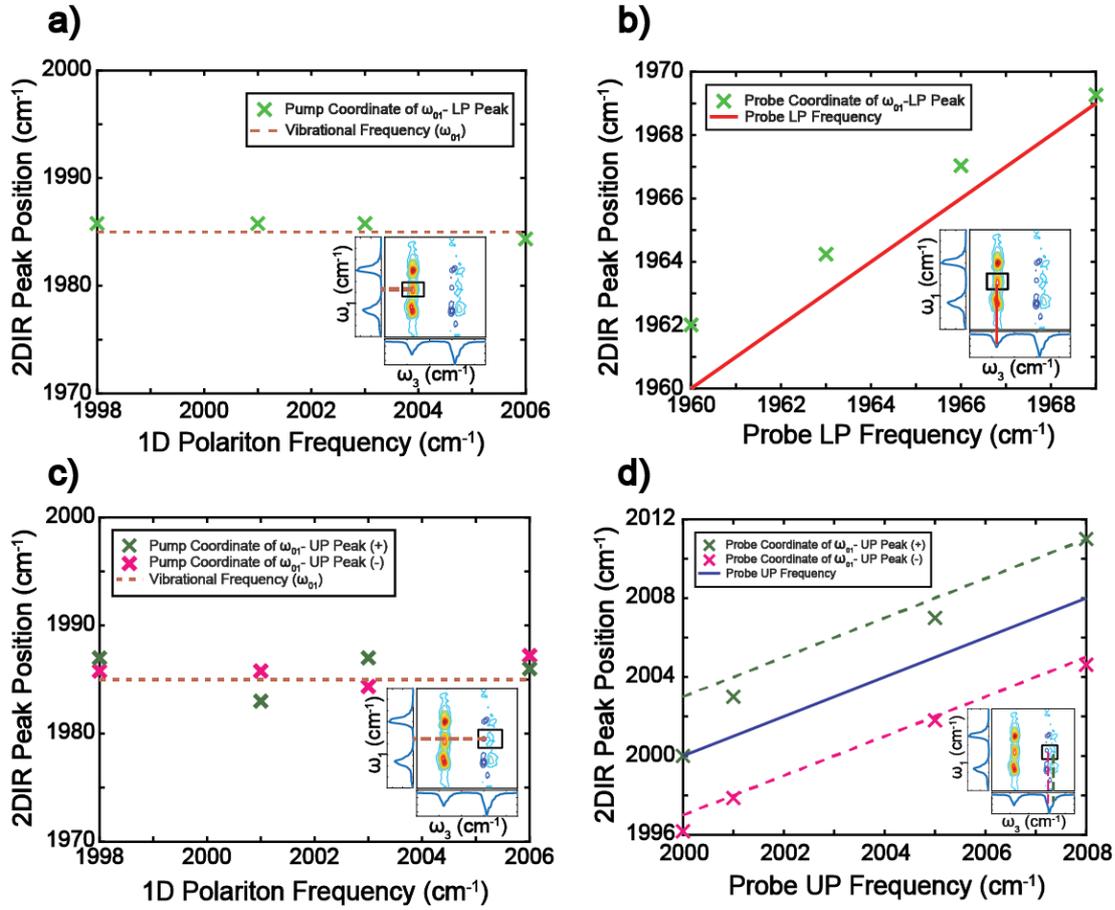

Figure S15. *(a)* $\omega_{01}$-*LP peak:* 2D spectral peak position projection onto $\omega_1$ axis versus UP peak frequency in 1D pump transmission spectrum; *(b)* $\omega_{01}$-*LP peak:* 2D spectral peak position projection onto $\omega_3$ axis versus LP peak frequency in 1D probe transmission spectrum; *(c)* $\omega_{01}$-*UP peak:* 2D spectral peak position projection onto $\omega_1$ axis versus UP peak frequency in 1D pump transmission spectrum; *(d)* $\omega_{01}$-*UP peak:* 2D spectral peak position projection onto $\omega_3$ axis versus UP peak frequency in 1D probe transmission spectrum.

In the case of the $\omega_{01}$-UP peaks, resonances associated to pumping $\omega_{01}$ and probing UP do exist, and are the only source of $\omega_{01}$-UP cross peaks. There are two reasons for this conclusion: 1) the shape of the spectral response along the probe axis of the $\omega_{01}$-UP peak indicates a strong signature of Rabi splitting reduction, as predicted in Secs. 3 and 4) no dark state-only response would appear at such high frequencies (see Fig. S12b). In consequence, we take the $\omega_{01}$-UP peak as direct evidence to support the fact that pumping dark states influence transitions of polaritons, as predicted by the theories presented in Secs. 3 and 4.

### (f) LP and UP Diagonal Peaks

Because the reservoir hot-band transition ($\omega_{12}$) is almost degenerate with respect to LP (when the detuning is small enough), we carefully analyze the origin of the LP diagonal peak. Based on Fig. S6a and b, the position of the LP diagonal peaks in the 2D IR spectra also roughly follow a linear relationship with respect to the position of the 1D LP transmission resonance, except for deviations in the projected position onto the pump axis of the 2D peaks when LP is close to either $\omega_{12}$ (1968 cm$^{-1}$) or $\omega_{23}$ (1952 cm$^{-1}$). This is like what we discussed in section SM5c: when the LP frequency is closer to those of dark modes, the latter blend in and influence the lineshape and intensity of the LP diagonal peak. Mainly, the interaction of LP-LP with the $\omega_{12}$-LP peak at a higher frequency (but same sign) is the reason for the blue shift. Similarly, it is likely true when the LP approaches the $\omega_{23}$ frequency, in which case the $\omega_{23}$ diagonal peak with opposite sign relative to the LP diagonal peak would cause the shift of the measured peak towards the blue. Thus, there exists clear contribution of dark modes in the LP diagonal region. Similar to the case of the UP-LP peak, we conclude the LP-LP peak has two contributions due to the (1) pumping and probing polaritons, and (2) pumping polaritons and probing hot band of the reservoir modes. As our theory shows (Sec. 4) these trends may also be understood in terms of the substantial hybridization of LP with bare-molecule hot band transitions due to their near-resonant character.

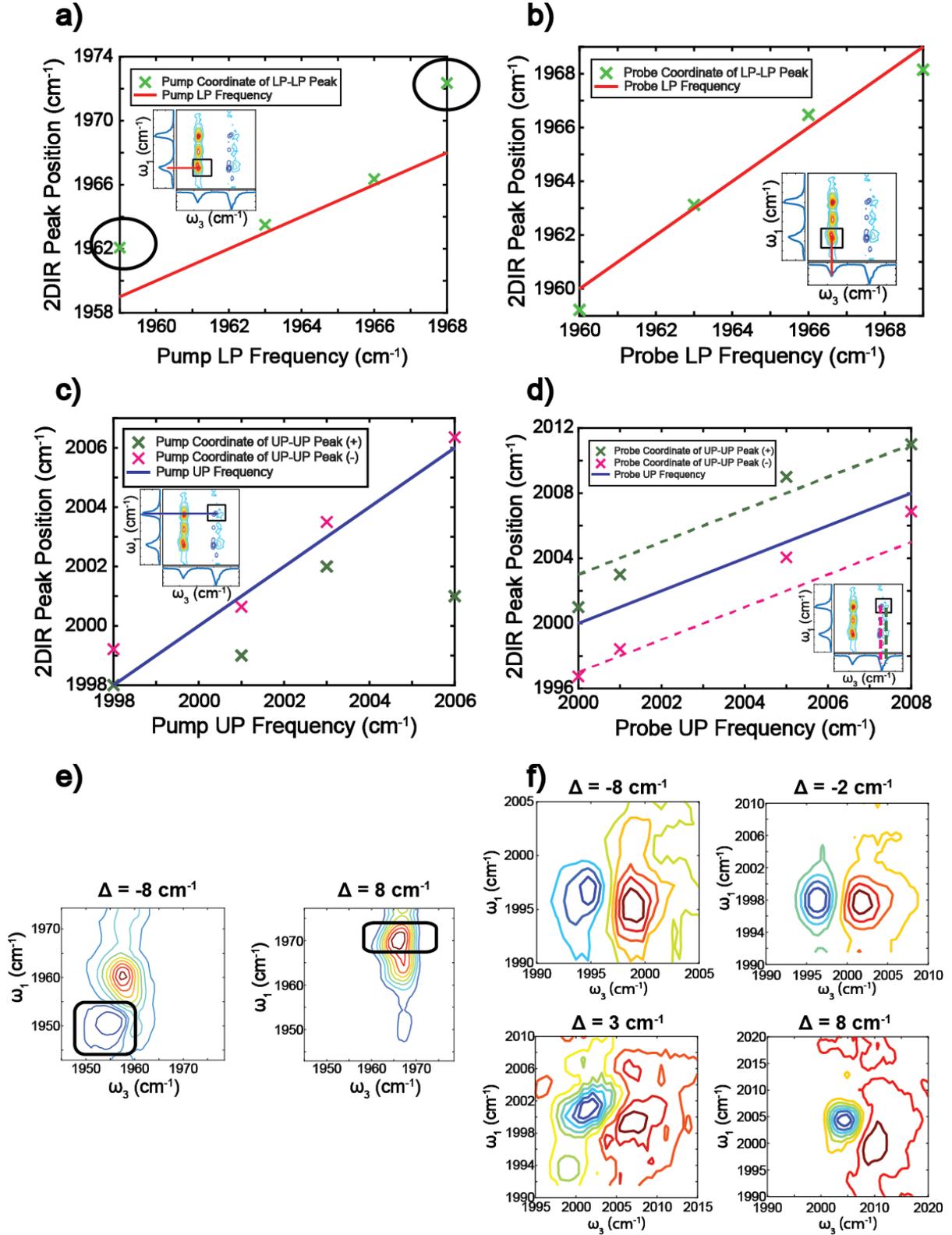

Figure S16. *(a)* LP diagonal peak: 2D spectral peak position projection onto $\omega_1$ axis versus LP peak frequency in 1D pump transmission spectrum; *(b)* LP diagonal peak: 2D spectral peak position projection onto $\omega_3$ axis versus LP peak frequency in 1D probe transmission spectrum; *(c)* UP diagonal peak: 2D spectral peak position projection onto $\omega_1$ axis versus UP peak frequency in 1D pump transmission spectrum; *(d)* UP diagonal peak: 2D spectral peak position projection onto $\omega_3$ axis versus UP peak frequency in 1D probe transmission spectrum; *(e)* Zoom-

in of LP diagonal peak at different cavity detuning values; *(f)* Zoom-in of UP diagonal peak at different cavity detuning values.

The characterization of the UP diagonal peak is much easier. The UP diagonal resonance coordinates roughly match the pump and probe UP frequencies (Fig. S16c and S16d) and its peak intensity reaches a maximum when dark state response is minimized. Thus, this peak corresponds to pumping UP and probing UP transitions. One notable characteristic of the spectrum is that the lineshape of the UP diagonal peak evolves from homogeneous to inhomogeneous with varying detuning (figure S16f). The mechanism behind the lineshape change could be due that at cavity is detuned, UP become more photon-like, which picks up the inhomogeneity of the cavity modes. Further investigation is necessary to properly understand this phenomenon.

## References

1. Shim, S.-H. & Zanni, M. T. How to turn your pump-probe instrument into a multidimensional spectrometer: 2D IR and Vis spectroscopies via pulse shaping. *Phys. Chem. Chem. Phys.* **11,** 748–761 (2009).

2. Deng, H., Haug, H. & Yamamoto, Y. Exciton-polariton Bose-Einstein condensation. *Rev. Mod. Phys.* **82,** 1489 (2010).

3. Simpkins, B. S. *et al.* Spanning Strong to Weak Normal Mode Coupling between Vibrational and Fabry – Pe´rot Cavity Modes through Tuning of Vibrational Absorption Strength. *ACS Photonics* **2,** 1460 (2015).

4. Dunkelberger, A. D., Spann, B. T., Fears, K. P., Simpkins, B. S. & Owrutsky, J. C. Modified relaxation dynamics and coherent energy exchange in coupled vibration-cavity polaritons. *Nat. Commun.* **7,** 13504 (2016).

5. McCoy, A. B., Guasco, T. L., Leavitt, C. M., Olesen, S. G. & Johnson, M. A. Vibrational manifestations of strong non-Condon effects in the $H_3O^+ \cdot X_3$ (X = Ar, $N_2$, $CH_4$, $H_2O$) complexes: A possible explanation for the intensity in the 'association band' in the vibrational spectrum of water. *Phys. Chem. Chem. Phys.* **14,** 7205 (2012).

6. Carusotto, I. & Ciuti, C. Quantum fluids of light. *Rev. Mod. Phys.* **85,** 299–366 (2013).

7. Knoester, J. & Mukamel, S. Transient gratings, four-wave mixing and polariton effects in nonlinear optics. *Physics Reports* **205,** 1–58 (1991).

8. Ribeiro, R. F. *et al.* Theory for nonlinear spectroscopy of vibrational polaritons. **submitted,** 1–5 (2017).

9. Khalil, M., Demirdöven, N. & Tokmakoff, A. Coherent 2D IR Spectroscopy: Molecular Structure and Dynamics in Solution. *J. Phys. Chem. A* **107,** 5258–5279 (2003).

10. Takemura, N. *et al.* Two-dimensional Fourier transform spectroscopy of exciton-polaritons and their interactions. *Phys. Rev. B - Condens. Matter Mater. Phys.* **92,** 125415 (2015).

11. Ebbesen, T. Hybrid Light–Matter States in a Molecular and Material Science Perspective. *Acc. Chem. Res.* **49,** 2403 (2016).

12. Houdré, R., Stanley, R. P. & Ilegems, M. Vacuum-field Rabi splitting in the presence of inhomogeneous broadening: Resolution of a homogeneous linewidth in an inhomogeneously broadened system. *Phys. Rev. A* **53,** 2711–2715 (1996).

13. Lindberg, M. & Binder, R. Dark states in coherent semiconductor spectroscopy. *Phys. Rev. Lett.* **75,** 1403–1406 (1995).